\documentclass[sigconf]{acmart}

\def\BibTeX{{\rm B\kern-.05em{\sc i\kern-.025em b}\kern-.08emT\kern-.1667em\lower.7ex\hbox{E}\kern-.125emX}}

\usepackage{soul} 
\usepackage{graphicx}
\usepackage{balance}
\usepackage{tabularx}

\copyrightyear{2020}
\acmYear{2020}
\setcopyright{acmlicensed}\acmConference[UMAP '20]{Proceedings of the 28th ACM Conference on User Modeling, Adaptation and Personalization}{July 14--17, 2020}{Genoa, Italy}
\acmBooktitle{Proceedings of the 28th ACM Conference on User Modeling, Adaptation and Personalization (UMAP '20), July 14--17, 2020, Genoa, Italy}
\acmPrice{15.00}
\acmDOI{10.1145/3340631.3394845}
\acmISBN{978-1-4503-6861-2/20/07}

\begin{document}
\title{
Personalized Recommendation of PoIs to People with Autism}

\author{Noemi Mauro}
 \orcid{}
 \affiliation{%
   \institution{Computer Science Department University of Torino}
   \streetaddress{Corso Svizzera 185}
   \city{Torino} 
   \state{Italy} 
   \postcode{10149}
 }
\email{noemi.mauro@unito.it}

\author{Liliana Ardissono}
 \orcid{}
 \affiliation{%
   \institution{Computer Science Department University of Torino}
   \streetaddress{Corso Svizzera 185}
   \city{Torino} 
   \state{Italy} 
   \postcode{10149}
 }
\email{liliana.ardissono@unito.it}

\author{Federica Cena}
 \orcid{}
 \affiliation{%
   \institution{Computer Science Department University of Torino}
   \streetaddress{Corso Svizzera 185}
   \city{Torino} 
   \state{Italy} 
   \postcode{10149}
 }
\email{federica.cena@unito.it}

\renewcommand{\shortauthors}{}

\begin{abstract}
The suggestion of Points of Interest to people with Autism Spectrum Disorder (ASD) challenges recommender systems research because these users' perception of places is influenced by idiosyncratic sensory aversions which can mine their experience by causing stress and anxiety. Therefore, managing individual preferences is not enough to provide these people with suitable recommendations.
In order to address this issue, we propose a Top-N recommendation model that combines the user's idiosyncratic aversions with her/his preferences in a personalized way to suggest the most compatible and likable Points of Interest for her/him. We are interested in finding a user-specific balance of compatibility and interest within a recommendation model that integrates heterogeneous evaluation criteria to appropriately take these aspects into account.
We tested our model on both ASD and ``neurotypical'' people.
The evaluation results show that, on both groups, our model outperforms in accuracy and ranking capability the recommender systems based on item compatibility, on user preferences, or which integrate these two aspects by means of a uniform evaluation model.

\end{abstract}



\maketitle

\section{Introduction}
\label{sec:introduction}
The personalized suggestion of Points of Interest (PoIs) challenges the research on recommender systems \cite{Ricci-etal:11} because, in order to provide truly inclusive services, different factors have to be taken into account, which go farther than modeling user interests.
Specifically, when suggesting PoIs to people with Autism Spectrum Disorder (ASD), we must take into account at least two aspects:

\begin{enumerate}
    \item 
The recommender system must work under data scarcity. There is a low number of users who can be analyzed to learn their interests: research indicates that Autism Spectrum Disorder affects around 1 in 100 people in EU \cite{elsabbagh2012global}. Moreover, ASD people are hard to be contacted because they have interaction problems and a tendency to avoid new experiences \cite{hobson1995}. Finally, their attention problems cause difficulties in providing detailed feedback about items \cite{murray2005attention}. 
\item
User preference management is not enough to generate useful suggestions. In fact, ASD people have idiosyncratic sensory aversions that influence the way they perceive items, especially places \cite{tavassoli2014sensory,robertson2017sensory, american2013diagnostic}. Therefore, traditional data about interests, used in recommenders for “neurotypical” individuals (i.e., not belonging to the autism spectrum), should be combined with these aversions because what bothers autistic people has great importance in their daily choices and can determine a high level of stress and anxiety \cite{gillott2007levels}. 
In order to take spatial needs into account in PoI recommendation, idiosyncratic aversions to noise, brightness and other sensory features have to be analyzed to recommend places that the user can perceive as safe and thus serenely experience. For ``safe PoIs'' we mean places which present ``safe'' characteristics from the sensory point of view; e.g., being quiet, scarcely crowded, or with smooth lights.
Notice that the inclusive recommendation of items generally goes beyond user preferences management: for example, in technology-enhanced learning, it is necessary to consider specific user features, such as learning capability, and corresponding specific item features, e.g., readability level \cite{pera2014automating}.
\end{enumerate}
Starting from Multi Criteria Decision Analysis \cite{vonWinterfeldt:86}, which provides techniques for the evaluation of multiple dimensions of items, and on match-making models based on user-to-item similarity \cite{Bridge-etal:05,Lops-etal:11}, most recommender systems, including collaborative multi-criteria ones \cite{Adomavicius-Kwon:07,Jannach-etal:14}, assume that the attributes of an item contribute to its utility to the user in an additive way. However, we notice that, depending on a person's idiosyncrasies and on their strength, problematic features might make items unsuitable for her/him even though they meet her/his preferences. Moreover, the impact of compatibility on users' choices varies individually and it cannot be separately managed with respect to preferences; e.g., some autistic people are determined to visit noisy and crowded places if they like them very much. Therefore, the models of item evaluation must reflect individual evaluation criteria by balancing feature compatibility and preference satisfaction.
In this work, we investigate the role of these two aspects in rating estimation, considering both ASD and neurotypical people. Specifically, we pose the following research questions:
\begin{itemize}
    \item
    {\em RQ1: in PoI recommendation, does a customized model of item evaluation, which balances feature compatibility and preference satisfaction in a personalized way, outperform recommender systems that manage only one of these aspects?}
    \item
    {\em RQ2: in PoI recommendation, does a customized model of item evaluation, which balances feature compatibility and preference satisfaction in a personalized way, outperform recommender systems which deal with both aspects but uniformly manage them?}
\end{itemize}
In order to answer these questions, we propose a novel Top-N recommender system that applies heterogeneous evaluation criteria to take user preferences and compatibility requirements into account by exploiting feature-based user profiles for the specification of individual needs.
Our recommender is focused on PoI suggestion to people with autism; however, it might be interesting to investigate its adaptation to other needs, e.g., related to motor disabilities, by extending the type of features which influence item compatibility. Our work has the following key aspects:
\begin{itemize}
    \item 
    We acquire data about people's aversion to sensory features in terms of disturbance caused by low or high feature values; e.g., darkness or strong light in physical places.
    For this purpose we use a questionnaire derived from \cite{tavassoli_sensoriality} but composed of a lower number of questions than that work.
    \item
    As our questionnaire provides data about users' aversion to a subset of the values that each feature can take, we also define general functions to estimate aversion to the whole range of possible values of the feature. Then, we derive feature compatibility with the user as the complement of aversion.
    \item
    For the estimation of item ratings, we distinguish user preferences for broad item categories from idiosyncratic sensory aversions. Moreover, as users might balance differently these aspects in item evaluation, we combine preferences and features compatibility by applying user-specific weights, which we acquire by analyzing users' ratings in conjunction with their declared preferences and idiosyncrasies. 
\end{itemize}
We tested our model on 20 adults with autism and 128 neurotypical ones\footnote{We had no mean to know whether the subjects of this second group belong to the spectrum or not.  However, we can reasonably expect that the sample respects the proportion of the entire population; thus, it might include 1 or 2 ASD persons at most (over 128). Henceforth, for simplicity, we refer to this sample as neurotypical people.}. On both groups of people, our model outperforms in accuracy and ranking capability a set of baseline recommender systems which singularly take item compatibility, or user preferences into account, and baselines that uniformly manage compatibility and preference information without differentiating their contribution. 
In summary, we provide the following novel contributions:
\begin{itemize}
    \item
    A Top-N PoI recommender system that fuses compatibility and preference data in rating estimation by balancing the impact of these aspects in a user-specific way.
    \item 
    A validation of the recommender on autistic and neurotypical people aimed at evaluating the performance of our model on both groups of users.
\end{itemize}
The approach presented in this paper is part of a wider ongoing project, PIUMA (Personalized Interactive Urban Maps for Autism)\footnote{PIUMA project involves a collaboration among the Computer Science and the Psychology Departments of the University of Torino and the Adult Autism Center of the city of Torino, Italy.}, which has the aim to develop novel digital solutions for helping people with ASD in their everyday movements \cite{DBLP:conf/chi/RappCBACBTKCB17, DBLP:conf/mhci/RappCMBSKB19}. The final result of PIUMA will be a mobile app showing maps customized to ASD users.
By means of the present work, we aim at adding a personalized selection of PoIs.  

In the following, Section \ref{sec:spatialNeeds} discusses the spatial needs of people with autism. Section \ref{sec:related} presents the related work.
Sections \ref{sec:knowledge} outlines how we retrieve information about PoIs, users' preferences and idiosyncrasies. Section \ref{sec:model} presents our model. Section \ref{sec:validation} describes the validation methodology we applied and Section \ref{sec:results} discusses the evaluation results. Section \ref{sec:conclusions} concludes the paper.

\section{Spatial needs of people with autism}
\label{sec:spatialNeeds}
Symptoms of autism span from severe language and intellectual disabilities in individuals with low- or mid-functioning autism, to no disabilities and an Intelligence Quotient (IQ) above the average in persons with high-functioning autism and Asperger’s syndrome.
Autism entails an atypical social functioning, which often results in avoiding everyday interactions \cite{hobson1995}. 
Questionnaire-based studies suggest atypical sensory perception in over 90\% of individuals with autism spectrum conditions  \cite{robertson2017sensory, golan2010enhancing, robertson2013relationship}, which means that individuals with autism appear  to react differently to sensory stimulations: a majority of them may become overwhelmed by environmental features that are easily managed by neurotypical people. For ASD individuals, the brain seems unable to appropriately balance the senses \cite{robertson2013relationship}. 
At least in part because of these characteristics, persons with ASD tend to have a reduced range of activities and interests, often preferring mechanical, deterministic situations, having the need to find reassurance by sticking to rigid, repetitious routines \cite{simm2016anxiety}.  It seems that they are less likely to explore new environments, and more likely to revisit well-known locations than neurotypical individuals \cite{smith2015spatial}. These peculiarities may entail idiosyncratic modes of perceiving space \cite{rapp2018designing}. 


Several studies  \cite{golan2010enhancing, robertson2013relationship, rapp2019spatial} reported that people with autism actively avoid places that may negatively impact on their senses.
Sight, smell and hearing are relevant in reference to mobility in urban environments and high sensory stimulation negatively influences individuals in their movements.  Further relevant environmental dimensions that could impact their sense of safeness are the temperature, openness, and crowding of a place. Such idiosyncratic sensory aversions may result in anxiety, fatigue, disgust, sense of oppression or distraction \cite{rapp2019spatial}. 

This should result in technological supports able to satisfy autistic people's idiosyncratic spatial needs, focusing on aversions derived from their high sensitivity to sensory stimulation. Moreover, there is a high need to personalize solutions because sensory sensitivity seems highly idiosyncratic; thus there are no features of places that may reassure the entire autistic population, and specific characteristics of each person should be considered \cite{putnam2019interactive}.

\section{Related work}
\label{sec:related}
 Technology is widely used to support people with autism in managing specific problems because they commonly exhibit an affinity with it \cite{putnam2008software,ramdoss2012computer}. In general, the research on autism tends to pay more attention to children \cite{goldsmith2004use} and it overlooks adults’ needs. This might be a consequence of the “medical model”, which promotes intervention toward school-aged individuals. Moreover, specifically the HCI community prefers to address social interaction problems \cite{putnam2019interactive,kientz2013interactive,grynszpan2014innovative}, likely because these are seen as the core characteristics of autism from a clinical point of view: therefore, that research focuses, e.g., on face-to-face conversation \cite{boyd2016saywat} and emotion management \cite{simm2016anxiety}, ignoring spatial difficulties. 

Most of the applications investigating the adoption of personalization strategies for ASD people regard the educational domain. 
For example, Judy et al. \cite{judy2012} present a personalized e-learning system that provides learning paths of different difficulty based on the user's past results.  They use ontologies to describe learning materials, annotation schemas and service ontologies, and they use a genetic algorithm as an optimization technique, representing a set of learning objects as chromosomes. No evaluation is provided. 

Garc{\'{\i}}a et al. \cite{DBLP:conf/ht/GarciaSFBFP16} propose an adaptive web-based application that helps students with autism overcome the challenges they may face when going to university. The adaptation consists of how the information site presents itself to autistic and non-autistic students but  the  information is the same for everyone. The adaptive functionality is based on learning styles (visual vs. verbal, global vs. analytical, active vs. reflective) and user history. For example, if the user is more visual than verbal, the video version of the content will be shown at the top of the learning object. Otherwise it will be moved to the bottom. No evaluation with ASD users is presented. 

Hong et al. \cite{hong2012designing} propose to provide autistic users with suggestions within a social network aimed at supporting the independence of young adults. However, they focus on the organization of the social network, by relying on peers' suggestions, instead of automatically generating recommendations to users. 

Differently, Costa et al. \cite{costa2017task} develop a task recommendation system that uses a case-based reasoning machine learning technique to supplement the child’s regular therapy.  The recommender suggests the daily activity to be performed (related to eating, keeping clean, getting dressed, etc.) based on  age, gender and time of day. It does not consider  the child's preferences, while the level of difficulty of the activities is manually set by the therapist. No evaluation with ASD people is described. 
Moreover, in \cite{Ng-Pera:18} Ng and Pera propose a hybrid game recommender for adult people with autism, based on collaborative and graph-based recommendation techniques, but they only carry out a preliminary test on neurotypical people.
Finally, in \cite{premasundari2019food} Premasundari and Yamini propose a system that recommends food and therapy for autistic children based on their symptoms. The system uses K-means algorithm for grouping the symptoms based on their type and association rule mining for the recommendation of food and therapies. However, the final target are the parents and caregivers, not people with autism. The system has been evaluated only from a usability point-of-view. 

Our work differs from the above ones for several reasons. Firstly, we focus on a different domain, i.e., spatial support. Secondly, we target adult people, who are the final users of the system and whose preferences and requirements have to be considered to succeed in item recommendation. Thirdly, we evaluated the approach with ASD people: this has rarely, if ever, been done in the related research.
Fourthly, our approach employs personal preferences for item categories and aversions to sensory features to steer recommendation in a context where a limited amount of feedback about items can practically be collected from users. This is different from the situation of other recommender systems in the health domain, which are targeted to users who are more willing to provide their preferences as needed, e.g., food recommendation \cite{Freyne-Berkovsky:10}.

Our work also differs from general content-based recommender systems \cite{Lops-etal:11}, feature-based \cite{Han-etal:05}, collaborative multi-criteria \cite{Adomavicius-Kwon:07,Jannach-etal:14,Zheng:17} and hybrid ones \cite{Burke:02,Gemmel-etal:12,Cantador-etal:11} because we treat sensory features of items as sources of discomfort for users rather than liking or disliking factors. We separately model the influence of idiosyncratic sensory features, which determine the compatibility of items with the user, from her/his preferences for types of items. Notice that this separation makes our model different from recommender systems that deal with negative preferences as well, e.g., \cite{Musto-etal:11}, because we support the management of heterogeneous criteria to deal with user preferences and sensory idiosyncrasies. 

Previously, the INTRIGUE \cite{Ardissono-etal:03} tourist guide introduced the notion of compatibility requirements in PoI recommendation but it did not investigate their different meaning and impact in the evaluation of items with respect to preferences. 

It is worth mentioning that, while constraint-based recommender systems \cite{Zanker-etal:10,Felfernig-etal:11,Wibowo-etal:18,Dragone-etal:18} are too knowledge intensive for our purposes (we are not suggesting item bundles with constraint satisfaction requirements), the optimization of soft constraints for path-finding under suitability criteria is relevant to our future work, in order to extend PoI recommendation with instructions for reaching the target places. This type of technique has been explored in recommender systems for routing; e.g., 
see \cite{Quercia-etal:14} and 
\cite{Verma-etal:18}.

\section{Preliminary study setup}
\label{sec:knowledge}
In this section we present how we gather data about users and POIS to run our experiments, as well as the sample's features.
\subsection{Data}
In order to gather data about autistic users' preferences and sensory aversions, we decided to bootstrap the user profiles by explicitly eliciting this type of information from people. Thus, we defined a questionnaire that will also be proposed to the users in the registration phase of the PIUMA app (not yet fully developed) in order to integrate an initial preference acquisition with the possibility to rate PoIs while the user is visiting them. 
The information about sensory aversions is hard to obtain \cite{tavassoli2014sensory}: usually, very long and complex questionnaires have to be completed for this purpose \cite{robertson2017sensory}. 
Moreover, asking people with ASD for such data is challenging because they have difficulties in social interactions and they tend to avoid new experiences \cite{schopler1986social}.
Given our users' attention problems \cite{bara2001, murray2005attention} and considering the application context, which is not a clinical setting, we decided to avoid long and detailed surveys. Consequently, we carefully prepared with psychologists a smaller questionnaire able to capture such information in a shorter way. 
Users filled in the questionnaire, possibly in the presence of an operator (when needed), and they answered to the question using the [1, 5] Likert scale. The questionnaire is composed of two sections:
\begin{itemize} 
\item 
In the first one we elicit user preferences about categories of PoIs such as restaurants, parks, and so forth; see the left column of Table \ref{tab:questions1}. 
\item
In the second one we gather information about users' aversions to sensory features of PoIs (right column).
We defined the questions about aversions by adapting the Sensory Perception Quotient (SPQ) test by Tavassoli et al. \cite{tavassoli_sensoriality}, a standard sensory questionnaire for adults with and without autism that assesses basic sensory hyper- and hyposensitivity.\footnote{The standard test is part of the battery of assessment tests that the people  compile when they become patients of the Autistic Adult center. It is made of 92 items, that is too much to be proposed to users in the context of a recommender system. In our questionnaire, we extrapolated the questions from Tavassoli's test. Specifically, we adapted the questions based on what users said in previous interviews carried out during a participatory design session for the design of the crowdsourcing system \cite{DBLP:conf/mhci/RappCMBSKB19}. The adaptation of the test is out of the scope of this paper.}
For some features (brightness and space), we asked the user to evaluate two extreme conditions, i.e., low or high levels, assuming that the middle ones are less problematic than the other ones. 
In other cases (crowding, noise and smell) we only asked about her/his annoyance in relation to the highest level because usually the low levels of these features are neutral. 
\end{itemize}

After the questionnaire, we asked the users to evaluate 50 specific PoIs located in Torino city center (e.g., \textit{How much do you like Castle Square?}) in order to collect a dataset of ratings to test our model.  We used the same [1, 5] Likert scale as above, but the ``I don't know the place'' option was available as well.
 
Those PoIs have been taken from Maps4All \cite{Maps4All} and they are representatives of all the categories of places it defines. Notice that we used an ad hoc 
crowdsourcing platform, designed to gather sensory features of places \cite{DBLP:conf/mhci/RappCMBSKB19}, because open data made available by geo web sites like OpenStreetMap \cite{OpenStreetMap} do not contain the specific sensory information we need. Maps4all allows the collection of such data: for each place, the user can rate in the [1, 5] scale its sensory features, and in particular its level of i) brightness, ii) crowding, iii) noise, iv) smell, v) openness, and vi) temperature.  
These sensory features have been defined on the basis of an authors' user study findings \cite{rapp2019spatial} and state-of-art research \cite{robertson2013relationship, robertson2017sensory}. The user can also provide a global evaluation of the place. For each datum, the system returns the mean evaluation it collected.
Currently, the Maps4All platform has been populated by means of two experimental crowdsourcing sessions during two lessons at the Master degree in Social Innovation and ICT at University of Torino, in May and December 2019. About 120 students have been involved in the crowdsourcing tasks, and they have been asked to provide evaluations for at least three PoIs each in Torino city centre. 
In total, during the two crowdsourcing sessions we collected the evaluations of 282 PoIs. Specifically, the 50 PoIs we considered for our study have been evaluated by at least three people each.

\begin{table}[t]
\centering
\caption{Short questionnaire to elicit information about preferences and sensory idiosyncrasies (translated from Italian).}
\vspace{-10pt}
\resizebox{\columnwidth}{!}{%
\setlength{\tabcolsep}{6pt}
\begin{tabular}{p{0.5\columnwidth}p{0.5\columnwidth}}
\toprule

 \textit{From 1 (not at all) to 5 (very much), how much do you like doing the following activities?}
\begin{itemize}
    \item be in nature, go to parks, gardens, green areas, \dots 
    \item visit museums, exhibitions, cultural events 
    \item go to the cinema, theater, concerts 
    \item go to comic shops 
    \item go to clothing stores 
    \item go to malls and markets 
    \item go to the library 
    \item go to the bookshop
    \item play sport  
    \item go to pubs, cafe
    \item go to the restaurant 
    \item go to the ice cream shop 
    \item stay in squares 
    \item go to the railway stations  
\end{itemize}
 &
 \textit{In a place, how much does it bother you:}
\begin{itemize}
    \item too much light  
    \item very low light
    \item a lot of people    
    \item a lot of noise 
    \item strong smells 
    \item cramped places (narrow, small) 
    \item large places
    \end{itemize}    
                      \\
\bottomrule
\end{tabular}}
\label{tab:questions1}
\end{table}

\subsection{Sample}
For our study we involved two groups of users\footnote{All participants signed a privacy consensus according to GDPR. University research ethical committee approval was obtained for the study}: 
\begin{itemize} 
\item 
20 ASD adults (from 22 to 40 years-old, mean age: 26,3, median 28; 11 men, 9 women) patients of the Autistic Adult center, medium- and high-functioning.
\item 
128 neurotypical subjects (from 19 to 71 years-old, mean age: 28,1, median 23; 63 men, 65 women) from university students and authors' contacts. 
\end{itemize}
Given the 50 PoIs selected, the mean number of PoI evaluations we obtained is 31 fpr ASD participants and 39 for neurotypical ones.

\section{Recommendation Model}
\label{sec:model}
As previously discussed, we assume that both user preferences and item compatibility should be taken into account to identify the most relevant items that an individual user can safely experience and like. However, evaluation criteria might be personal and users could weight these aspects differently in their decision-making processes. For instance, in contrast to the tendency of people with ASD to visit places in which they feel comfortable, during our participatory design interviews sessions, we had the chance to  interact with a guy with autism who frequently visits bowling halls, regardless of the negative impact of noise and crowd on his senses, because he likes bowling so much that he does not want to give it up. We thus propose a recommendation model which, based on the observed item evaluations, can weight the contribution of compatibility and preferences in rating prediction on a user-specific basis. For clarity purposes we split the description of our model as follows:
\begin{enumerate}
\item 
input data for recommendation (Section \ref{sec:input});
\item 
estimation of the compatibility of the individual features of an item with the user (Section \ref{sec:featureCompatibility});
\item 
estimation of the overall compatibility of the item with the user (Section \ref{sec:itemCompatibility});
\item  
preference-based item evaluation (Section \ref{sec:preferences});
\item 
integration of compatibility and preference-based evaluation to predict the user's rating of the item (Section \ref{sec:overallItemEvaluation}).
\end{enumerate}
\begin{figure}[t]
    \centering
    \includegraphics[width=1.0\linewidth]{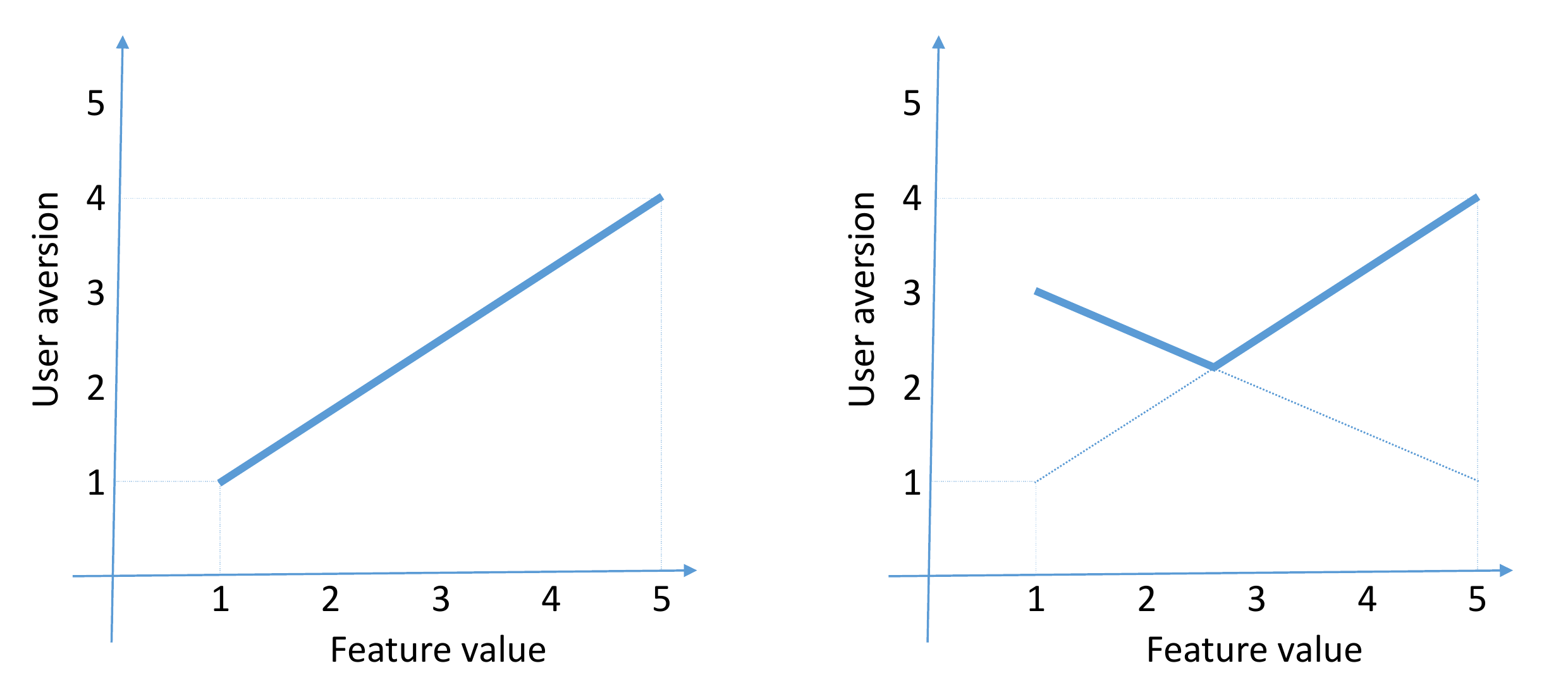}
    \caption{Representation of a user's aversion to a feature.}
    \label{fig:aversion}
\end{figure}
Before describing our model we introduce the notation we use:
\begin{itemize} 
    \item
    $U$ is the set of users and $I$ the set of items of the domain. 
    \item 
    $C$ is the set of item categories; e.g., shops, cinemas, etc..
    \item 
    $L$ is a Likert scale in $[1, v_{max}]$; in this work, $v_{max}=5$.
    \item 
    $F = F^\uparrow \cup F^V$ is the set of sensory features defined in our domain. We assume that each feature $f \in F$ takes values in $L$. Specifically:
    \begin{itemize}
        \item 
        $ F^\uparrow$ is the set of features $f$ such that, the higher the value of $f$, the stronger its negative impact on the user; e.g., noise. 
        \item 
        $F^V$ denotes features whose extreme values make users uncomfortable, while the middle ones are less problematic; e.g., brightness. 
    \end{itemize}
    In our domain there are no features such that people are expected to feel comfortable with high values and uncomfortable with low ones; therefore, we omit this class.
\end{itemize}
For each user $u \in U$ and item $i \in I$, we estimate $u$'s evaluation of $i$ (denoted as $\hat{r}_{ui}$) as a decimal number in the [1, $v_{max}$] interval, by taking $u$'s previous ratings, preferences for item categories and idiosyncrasies into account.

\subsection{Input Data}
\label{sec:input}
Our model takes two types of information as input:
\begin{itemize}
    \item 
    The profile of $u \in U$, extracted from the questionnaire  
    data, which specifies:
    \begin{itemize} 
        \item 
        The ratings $r_j$ in $L$ (s)he provided for a set of items $j \in I$. 
        \item 
        Her/his declared preferences for the categories $c \in C$, each one expressed in the $L$ scale.
        \item 
        Her/his declared sensory aversion to specific values of item features, expressed in $L$. We denote $u$'s aversion to a value $v$ of a feature $f \in F$ as $a_{ufv}$. For example, $a_{uf5} = 4$ means that $u$ is fairly disturbed by items having $f=5$. 
        \begin{itemize}
            \item 
            For each feature $f \in  F^\uparrow$ we assume by default that $a_{uf1} = 1$. Therefore, the user profile stores a single value, $a_{ufv_{max}}$, which specifies $u$'s aversion to the maximum value of $f$; i.e., $v_{max}$.
            \item
            For each feature $f$ in $F^V$, the user profile stores two values which express $u$'s aversion to the minimum and maximum values of $f$ respectively: e.g., \{$a_{uf1} = 3$, $a_{ufv_{max}} = 4$\}. 
        \end{itemize}
    \end{itemize}
    \item 
    Item $i$'s profile, extracted from the crowdsourced data, which specifies:
    \begin{itemize}
    \item 
    The category $c \in C$ of $i$.
    \item 
    A vector $\vec{\bf {i}}$ storing, for each feature $f \in F$, the value of $f$ in item $i$ (denoted as  $\vec{{\bf i}}_f^{\,}$) retrieved by querying Maps4All; $\vec{{\bf i}}_f^{\,}$ takes values in the [1, $v_{max}$] interval.
    \end{itemize}
\end{itemize}

\subsection{Compatibility of Individual Features with the User}
\label{sec:featureCompatibility}
We can define compatibility as the opposite of aversion in the range of values that features can take. However, user profiles only include one or two aversion values declared by users for each feature and the missing ones have to be interpolated.
We thus define two patterns to approximate users' idiosyncratic aversions to item features starting from the values stored in the user profile: 
\begin{itemize}
    \item
    For each $f \in F^\uparrow$ we approximate aversion as a linearly increasing function. If we represent feature values in the $X$ axis and user aversion in the $Y$ axis of a plane, we can define this function as a line which connects point (1, 1) to point ($v_{max}$, $a_{ufv_{max}}$), as in the left graph in Figure \ref{fig:aversion}:
        \begin{equation}
        line^\uparrow(x) = 1 + \frac{(a_{ufv_{max}}-1)(x-1)}{v_{max}-1}
        \label{eq:crescente}
    \end{equation}
    We thus estimate $u$'s aversion to $f$ in $i$ ($ea_{ufi}$) as follows:
    \begin{equation}
        ea_{ufi} = line^\uparrow(\vec{{\bf i}}_f^{\,})
        \label{eq:upLine}
    \end{equation}
    \item 
    As far as $F^V$ is concerned, and given \{$a_{uf1}, a_{ufv_{max}}$\} in $u$'s profile, we approximate aversion by means of a concave function on the range of $f$. The aversion function has a "V" shape which we approximate by drawing two lines, as the right graph of Figure \ref{fig:aversion}: 
    \begin{itemize} 
        \item 
        The first line ($line^\uparrow$) connects points (1, 1) and ($v_{max}$, $a_{ufv_{max}}$) to represent the increment of aversion towards the maximum value of $f$. 
        \item 
        The second line ($line_\downarrow$) connects points (1, $a_{uf1}$) and ($v_{max}$, 1) to represent the decrease in aversion while $f$ takes higher values than its minimum:
        \begin{equation}
        line_\downarrow(x) = 1+\frac{(x-v_{max})(1 - a_{ufv_{min}})}{v_{max}-1}
        \label{eq:lineDown}
    \end{equation}
    \end{itemize}
    Therefore, we estimate $u$'s aversion to $f$ in $i$ by selecting the maximum values of the two lines, i.e.:
    \begin{equation}
        ea_{ufi} = \max (line^\uparrow(\vec{{\bf i}}_f^{\,}), line_\downarrow(\vec{{\bf i}}_f^{\,}))
        \label{eq:v}
    \end{equation}
    This corresponds to the broken thick line in Figure \ref{fig:aversion}.
\end{itemize}
Notice that $ea_{ufi}$ takes values in the [1, $v_{max}$] interval and higher values of this measure mean that the feature generates more discomfort to $u$.
Given $ea_{ufi}$, the compatibility of $f$ with $u$ in $i$, denoted as $comp_{fui}$ (in [1, $v_{max}$]), can thus be defined as:
\begin{equation}
    comp_{fui} = v_{max} + 1 - ea_{ufi}
\end{equation}

\subsection{Overall Item Compatibility: Aggregation Measures}
\label{sec:itemCompatibility}
We propose alternative aggregation measures to compute the overall compatibility of an item $i$ with a user $u$ ($comp_{iu}$) by modeling different types of influence of individual features. In Section \ref{sec:results} we evaluate their performance, in combination with different recommendation algorithms.
\begin{itemize}
    \item
    {\bf Min:} $comp_{iu}$ is the minimum compatibility of $i$'s features with $u$:
    \begin{equation}
    \label{eq:min}
        comp_{iu} = \min\limits_{f\in F} comp_{fui}
    \end{equation}
    This measure is conjunctive: it evaluates an item as incompatible with $u$ if it has at least one totally incompatible feature.
\end{itemize} 
\begin{itemize}
    \item
    {\bf Ave:} $comp_{iu}$ is the mean compatibility of $i$'s features:
    \begin{equation}
    \label{eq:mean}
        comp_{iu} = \frac{\sum\limits_{f\in F} comp_{fui}}{|F|}
    \end{equation}
    where $| \cdot |$ denotes set cardinality. This measure is additive (disjunctive) and equally balances the influence of all the features on compatibility.
\end{itemize} 
The following aggregation measures estimate overall compatibility in function of the distance between the features of $i$ (stored in the $\vec{{\bf i}}$ vector) and those of an $ideal$ item which best matches $u$'s idiosyncrasies ($\overrightarrow{{\bf{ideal_u}}}$). For each $f \in F$, $\overrightarrow{{\bf{ideal_{uf}}}}$ is the most compatible value of $f$ on the basis of $u$'s estimated aversion presented in Section \ref{sec:featureCompatibility}. This value is $1$ for $f \in F^\uparrow$, while it is the feature value associated to the minimum aversion for $f \in F^V$; e.g., $\overrightarrow{{\bf{ideal_{uf}}}} = 2.7$ in the right graphic of Figure \ref{fig:aversion} ($X$ axis). The overall compatibility of $i$ with $u$ is thus computed as follows:
\begin{itemize}
    \item
    {\bf Cos:} $comp_{iu}$ is the Cosine similarity between $\vec{{\bf i}}$ and $\overrightarrow{{\bf{ideal_u}}}$:
    \begin{equation}
    \label{eq:cosine}
        comp_{iu} = \frac{\vec{{\bf i}} \cdot \overrightarrow{{\bf{ideal_u}}}} {\Vert \vec{{\bf i}}\Vert_F \; \Vert \overrightarrow{{\bf{ideal_u}}}\Vert_F}
    \end{equation}
    where $\cdot$ is the scalar vector product and $\Vert \cdot \Vert_F$ is the Frobenius Norm.
    The smaller is the angle between $\vec{{\bf i}}$ and $\overrightarrow{{\bf{ideal_u}}}$, the more compatible is $i$ with $u$. 
    \item
    {\bf RMSD:} $comp_{iu}$ is the complement of the Root Mean Square Deviation between $\vec{{\bf i}}$ and $\overrightarrow{{\bf{ideal_u}}}$:
    \begin{equation}
    \label{eq:rmsd}
        comp_{iu} = v_{max} +1 - \sqrt{\frac{1} {|F|}* \sum\limits_{f\in F} ( \vec{{\bf i}}_f  - \overrightarrow{{\bf ideal}_{uf}} )^2}
    \end{equation}
    The smaller is the distance between $\vec{{\bf i}}$ and $\overrightarrow{{\bf{ideal_u}}}$, the more compatible is $i$ with $u$. 
\end{itemize}

\subsection{Preference-based Item Evaluation}
\label{sec:preferences}
While compatibility indicates whether the user can safely experience an item, it does not mean that (s)he likes it. User preferences have to be taken into account for this purpose.

In our domain, the only preference that we consider is the user's interest in the category of the item to be evaluated. Thus, the preference value of a user $u$ for an item of category $c \in C$ corresponds to the value of $u$'s preference for $c$ stored in $u$'s profile. We denote this value as $p_{uc}$.

It is worth mentioning that, if more preferences had to be modeled, a classical Multi-Criteria Decision Analysis approach might be applied to compute an overall preference estimation as a weighted function of preferences for individual attributes \cite{Adomavicius-Kwon:07,Adomavicius-Kwon:11,Jannach-etal:14}. However, this is out of the scope of the present work. 

\subsection{Rating Prediction}
\label{sec:overallItemEvaluation}
In order to take personal balance between compatibility and preferences into account, we propose to identify user-dependent evaluation criteria by exploiting the user's idiosyncrasies and preferences in combination with the ratings of items (s)he provides, which we consider as the ground-truth revealing her/his priorities.
For this purpose, we define the overall evaluation of items, which produces the estimated rating for each user, as a weighted mean of items compatibility and user preferences:
\begin{equation}
    \hat{r}_{ui} = \alpha * comp_{iu} + (1-\alpha) * p_{uc_i}
    \label{eq:est_rating}
\end{equation}
where $\alpha$ takes values in the [0, 1] interval and $p_{uc_i} \in L$ is the preference-based evaluation of $i$, given $u$'s profile. This model, henceforth referred as {\bf Ind} (i.e., Individual), identifies a specific $\alpha$ value for each user to optimize item recommendation to her/him. 
We identify the value of $\alpha$ for each $u \in U$ as the one that minimizes the distance between estimated ratings and ground-truth ones. 

\section{Validation methodology}
\label{sec:validation}
We aim at assessing the usefulness of modeling both compatibility and preference aspects for recommendation, with respect to taking only one of these aspects into account. Moreover, we aim at evaluating the usefulness of a personalized balance between these two types of information, specified by the $\alpha$ parameter of Equation \ref{eq:est_rating}. For these purposes we compare our model to baseline recommenders which (i) uniformly manage compatibility and user preferences, i.e., they ignore the possibly different impact on decision-making) or (ii) focus on a single aspect, either compatibility or preferences.
Specifically, we consider the following baselines:
\begin{itemize}
    \item 
    {\bf Multi-Criteria (MC):} this recommender system estimates item ratings by uniformly treating idiosyncratic features and preferences on the basis of the aggregation measures described in Section \ref{sec:itemCompatibility}. It computes $\hat{r}_{ui}$ by fusing $u$'s preference for the category of $i$ ($p_{uc_i}$) with the compatibility of individual features ($comp_{fui}$) by means of a single aggregation function; e.g., the mean of all these values as in Ave (Equation \ref{eq:mean}). 
    \item 
    {\bf C-only:} this is a setting of our recommendation model (Equation \ref{eq:est_rating}) in which $\alpha=1$ to evaluate items exclusively on the basis of their compatibility with the user.
    \item 
    {\bf Pref-only:} in this setting of our model, $\alpha=0$ to evaluate items on the exclusive basis of the user's preferences.
\end{itemize}
We did not select as baselines any collaborative or feature-based recommenders such as those proposed in \cite{Han-etal:05} or \cite{Adomavicius-Kwon:07} because there is not enough data to train the recommenders.

We compare our model to the above baselines on both the autistic users dataset (henceforth denoted as AUT) and on the neurotypical users one (NOR). For the comparison we configure all the algorithms on each aggregation measure of Section \ref{sec:itemCompatibility}. For each algorithm, we denote the specific configuration we apply by appending its name to that of the algorithm; e.g., Ind$_{Cos}$ represents the application of the Cos aggregation measure to model Ind.

To evaluate recommendation performance we focus on ranking capability (MRR and MAP), accuracy (Precision, Recall and F1), error minimization (MAE and RMSE) and user coverage.
We perform a 5-fold cross validation in which, for every fold, we use 80\% as training set and 20\% as test set. As the Ind models have to optimize the $\alpha$ parameter, we train each of them to find the best user-specific setting by optimizing its results with respect to MAP. Moreover, to be sure that the baselines are consistently evaluated, we run the other algorithms (MC, C-only and Pref-only, which do not need any training) on the same test sets used for Ind.

\section{Evaluation results and discussion}
\label{sec:results}
Tables \ref{tab:autistici} and \ref{tab:neurotipici} show the Top-N evaluation results with N=5; i.e., the list of suggested items has length=5. The tables omit the results concerning user coverage because it is 100\% in all the cases.

We consider two categories of algorithms, i.e., the configurations of our model on the various aggregation measures and the corresponding ones of the baselines, and we use the following notation:
the best value across all algorithms is printed in bold; the best value obtained by the other category of algorithms is underlined (when our model obtains the best value we underline the best value achieved by the baselines, and {\em vice versa}). Stars indicate significant differences according to a Student T-Test between the best performing algorithm from each category; **: p<0.01; *: p<0.05.
It can be noticed that:
\begin{itemize}
    \item 
    {\it Ind$_{Cos}$ excels in accuracy and ranking capabilities:} on both datasets it outperforms all the other algorithms (baselines and own category) in F1 and MAP, and it has the best Recall of its own category. Moreover, it obtains better MRR values than all algorithms on AUT and most baselines on NOR. As far as Precision is concerned, Ind$_{Cos}$ is the third best algorithm on AUT while it is the fourth best on NOR; in both cases it is outperformed by another member of its own category. {\it The results concerning error minimization are mixed:} on both datasets the MAE of Ind$_{Cos}$ is worse than that of most of the other algorithms, including  those of its own category. However, on NOR, it has the best RMSE of its own category and it achieves better results than most of the baselines.
    \item 
    {\it Ind$_{Min}$ excels in error minimization:} on both datasets it obtains the best MAE of all algorithms; moreover, concerning RMSE, it outperforms all algorithms in AUT and the other algorithms of its own category in NOR. {\it Ind$_{Min}$ also has fairly good ranking capability:} it is the second best algorithm for MAP on both datasets and it has the best MRR on NOR, and second best MRR of its own category on AUT. {\it It has good Precision:} it achieves the best one of all algorithms on AUT and the third best on NOR. On both datasets it outperforms most baselines in Recall. Moreover, on AUT it has the third best F1 of all algorithms (lower than Ind$_{Cos}$ and C-only$_{Cos}$), while its accuracy on NOR is in middle position. 
    \item
 
    {\it Ind$_{Ave}$ has the third best MAP  results of its own category, the third and second best MRR on AUT and NOR respectively, } and, concerning these metrics, it is positioned in the higher part of the general classification. As far as Precision, Recall and F1 are concerned, it achieves intermediate results on AUT but it is placed in the top positions of the classification on NOR. It has {\it fairly good error minimization capability (MAE, RMSE),} better than several baselines, and it is placed in second position within its own category on both datasets.
    
    \begin{table}[t]
\centering
\caption{Results on AUT dataset for N=5. The lines of the table are ordered by MAP. The best values of each measure across all algorithms is printed in bold. The best value obtained by the other category of algorithms is underlined. Stars denote statistical significance: **: p<0.01; *: p<0.05.}
\vspace{-10pt}
\resizebox{\columnwidth}{!}{%
{\def\arraystretch{1.4}

\begin{tabular}{lrrrrrrr}
\toprule
Algorithm       & \multicolumn{1}{c}{Prec.} & \multicolumn{1}{c}{Recall} & \multicolumn{1}{c}{F1} & \multicolumn{1}{c}{MAP} & \multicolumn{1}{c}{MRR} & \multicolumn{1}{c}{MAE} & \multicolumn{1}{c}{RMSE} \\ \midrule
Ind$_{Cos}$     & 0.6290             & \underline{0.6207} & \textbf{0.6046}    & \textbf{**0.5384}  & \textbf{0.8095}    & 0.9927             & 1.4541             \\
Ind$_{Min}$     & \textbf{0.6328}    & 0.5832             & 0.5910             & 0.5125             & 0.7825             & \textbf{0.8691}    & \textbf{*1.3020}   \\
Pref-only       & 0.6220             & 0.5912             & 0.5860             & \underline{0.5114} & 0.7858             & \underline{0.9346} & \underline{1.4276} \\
Ind$_{Ave}$     & 0.6118             & 0.5710             & 0.5736             & 0.4960             & 0.7667             & 0.9168             & 1.3659             \\
C-only$_{Cos}$  & 0.6263             & \textbf{0.6224}    & \underline{0.6001} & 0.4877             & 0.7583             & 1.3675             & 1.6948             \\
Ind$_{RMSD}$    & 0.5978             & 0.5545             & 0.5577             & 0.4799             & 0.7537             & 0.9965             & 1.4533             \\
MC$_{Ave}$      & 0.6255             & 0.5383             & 0.5575             & 0.4489             & 0.7792             & 1.1902             & 1.4861             \\
MC$_{RMSD}$     & 0.6080             & 0.5396             & 0.5463             & 0.4429             & 0.7775             & 1.2172             & 1.5426             \\
MC$_{Min}$      & \underline{0.6305} & 0.5057             & 0.5344             & 0.4352             & \underline{0.7950} & 1.4512             & 1.7943             \\
MC$_{Cos}$      & 0.5917             & 0.5558             & 0.5459             & 0.4336             & 0.7217             & 1.3534             & 1.6236             \\
C-only$_{Min}$  & 0.6065             & 0.4999             & 0.5230             & 0.4166             & 0.7583             & 1.3675             & 1.6816             \\
C-only$_{Ave}$  & 0.5912             & 0.5154             & 0.5270             & 0.4142             & 0.7192             & 1.3045             & 1.6060             \\
C-only$_{RMSD}$ & 0.5825             & 0.5009             & 0.5145             & 0.4036             & 0.7142             & 1.3702             & 1.7168             \\ \bottomrule
\end{tabular}}
}
\label{tab:autistici}
\end{table}

\begin{table}[t]
\centering
\caption{Results on NOR dataset for N=5. We use the same notation of Table \ref{tab:autistici}.}
\vspace{-10pt}
\resizebox{0.96\columnwidth}{!}{%
{\def\arraystretch{1.4}

\begin{tabular}{lccccccc}
\toprule
Algorithm       & \multicolumn{1}{c}{Prec.} & \multicolumn{1}{c}{Recall} & \multicolumn{1}{c}{F1} & \multicolumn{1}{c}{MAP} & \multicolumn{1}{c}{MRR} & \multicolumn{1}{c}{MAE} & \multicolumn{1}{c}{RMSE} \\ \midrule
Ind$_{Cos}$     & 0.5790             & \underline{0.5406} & \textbf{0.5349}    & \textbf{0.4139}    & 0.7475             & 1.1792             & 1.5232             \\
Ind$_{Min}$     & 0.5791             & 0.5225             & 0.5250             & 0.4120             & \textbf{0.7688}    & \textbf{1.0950}    & \underline{1.4024} \\
Ind$_{Ave}$     & 0.5740             & 0.5261             & 0.5262             & 0.4108             & 0.7555             & 1.1085             & 1.4343             \\
Ind$_{RMSD}$    & \textbf{0.5816}    & 0.5286             & 0.5297             & 0.4108             & 0.7521             & 1.1427             & 1.4758             \\
Pref-only       & \underline{0.5795} & 0.5408             & \underline{0.5347} & \underline{0.4076} & 0.7304             & 1.1416             & 1.5270             \\
C-only$_{Cos}$  & 0.5503             & \textbf{0.5414}    & 0.5255             & 0.4000             & 0.7189             & 1.4374             & 1.7456             \\
MC$_{Ave}$      & 0.5752             & 0.5154             & 0.5213             & 0.3995             & 0.7564             & \underline{1.1238} & {\bf 1.3564}       \\
MC$_{Min}$      & 0.5664             & 0.4956             & 0.5053             & 0.3890             & \underline{0.7583} & 1.1249             & 1.4052             \\
MC$_{RMSD}$     & 0.5568             & 0.4840             & 0.4963             & 0.3767             & 0.7433             & 1.3320             & 1.6255             \\
C-only$_{Ave}$  & 0.5476             & 0.4936             & 0.4979             & 0.3701             & 0.7168             & 1.2122             & 1.4668             \\
C-only$_{Min}$  & 0.5507             & 0.4769             & 0.4899             & 0.3673             & 0.7359             & 1.1704             & 1.4213             \\
C-only$_{RMSD}$ & 0.5460             & 0.4870             & 0.4936             & 0.3651             & 0.7223             & 1.4157             & 1.7281             \\
MC$_{Cos}$      & 0.5274             & 0.5053             & 0.4974             & 0.3535             & 0.6591             & 1.2775             & 1.5795             \\ 
 \bottomrule
\end{tabular}}}
\label{tab:neurotipici}
\vspace{-3mm}
\end{table}

    \item 
    {\it Ind$_{RMSD}$ is the worst performing configuration of our model:} on AUT it obtains the lowest performance results of its own category on all measures, but it still outperforms several baselines in MAP and in other metrics. Differently, in NOR it achieves fairly good F1, the best Precision of all algorithms (including baselines), fairly good MAP and MRR and fairly good error minimization.
    \item 
    {\it Pref-only is the best baseline regarding MAP} and it has {\it fairly good MRR and F1} on both datasets. This algorithm achieves the best MAE and RMSE of the baselines on AUT, while on NOR it is outperformed by other baselines. Except for RMSE in AUT and MAE in NOR, it achieves lower results than Ind$_{Cos}$ on all performance metrics.
    \item 
    On both datasets {\it C-only$_{Cos}$ has lower ranking capability than the previous algorithms (MAP and MRR) but it has fairly good accuracy,} being the best, or second best baseline on the various measures. Its error minimization capability is definitely lower than that of our model. Notice that the other configurations of C-only (i.e., using Min, Ave and RMSD) have worse performance than this one on both datasets, in all metrics except for error minimization.
    \item 
    Similarly, {\it the configurations of MC, except for MC$_{Min}$ in Precision and MRR, have middle to low performance}; they are outperformed by our model or by some other algorithms in both datasets.
\end{itemize}

The evaluation results suggest that Ind$_{Cos}$ is the best recommendation algorithm because it combines good ranking capability with good accuracy. Ind$_{Min}$ achieves better error minimization than this algorithm but, as previously discussed, this is a secondary evaluation criterion for us. 

Unfortunately, the low size of the AUT and NOR datasets does not support the statistical significance of results for several metrics. However, the results concerning MAP (and RMSE) on the AUT dataset are significant. This is important because our recommendation model is targeted to autistic people and we can thus rely on the ranking capability results we obtained on them. At the same time, the results are encouraging for neurotypical users. Thus, it is worth investigating performance within a larger experiment that will possibly provide more statistically relevant results on both groups of people.

The evaluation results help us answering our research questions:
\begin{itemize}
    \item 
    {\em RQ1:} we can positively answer this question. As far as F1 and ranking capability are concerned, the configurations of Ind that take both preferences and compatibility into account (and, specifically, Ind$_{Cos}$) outperform Pref-only, which only exploits user preferences to recommend items. Moreover, they achieve better results than all the algorithms that only use compatibility information (C-only). These algorithms are outpeformed by Pref-only, too. This means that, not surprisingly, compatibility information alone is not enough to generate relevant recommendations for the user.
    \item
    {\em RQ2:} we can positively answer this question as well. In fact, the Ind configurations outperform the MC ones, regardless of the used aggregation measure, in most evaluation metrics and especially in ranking capability and F1.
\end{itemize}
To summarize, in Top-N PoI recommendation preference information is useful to suggest relevant items. However, better results can be achieved by combining this type of information with a compatibility evaluation aimed at assessing whether the user can serenely experience the recommended items. Interestingly, a uniform management of compatibility and preference information, which does not distinguish the possibly heterogeneous evaluation criteria concerning them, does not bring good results. Conversely, the acquisition of user-specific weights to balance the impact of compatibility and interests in item evaluation improves item suggestion.

\section{Conclusions}
\label{sec:conclusions}
Users with Autism Spectrum Disorder (ASD) are a particularly interesting and challenging target of PoI recommender systems because of their characteristics and needs in relation to places. In order to suggest suitable solutions, which the user can like {\em and} serenely experience, both her/his preferences for PoI categories, traditionally analyzed by researchers, and her/his aversions to sensory features, have to be considered: as a matter of fact, the latter can seriously affect ASD people's experience of the places, causing negative feelings. Notice that it is particularly important to avoid wrong suggestions because there might be critical effects on the person, given her/his “frailty”, causing for example anxiety, irritation and anger, with unpredictable consequences.

In this paper we presented a Top-N recommender of PoIs for ASD people that takes their idiosyncratic aversions to sensory features into account in order to generate suggestions that are expected to be both pleasant and safe for them.
We tested our model on autistic and neurotypical users.
The evaluation results show that, on both user groups, our model outperforms in accuracy and ranking capability baseline recommenders which (i) evaluate items on the sole basis of how closely they meet the user's preferences, or how compatible they are with her/his idiosyncratic aversions to sensory features, and (ii) uniformly manage compatibility and preference information without distinguishing the possibly different contributions of these aspects to item evaluation.
We thus conclude that the integration of possibly heterogeneous evaluation criteria concerning user interests and idiosyncratic aversions is a promising approach to extend the adoption of recommender systems to new user groups with respect to those typically addressed in the state of the art.

Two main limitations of our approach concern the data about PoIs and users available for the experiment.
We plan to address these limitations as follows: 
\begin{itemize} 
\item 
 We will extend information about PoIs in two ways: (i) by starting a VGI campaign with people with ASD and their caregivers, as well as with the general population, to acquire a larger amount of data, and (ii) by extracting sensory information from consumer reviews available in online platforms such as TripAdvisor \cite{TripAdvisor}; e.g., see \cite{Bilici-Saygin:17}.
\item When the PIUMA app will be available, we will be able to learn detailed information about user interests and aversions by coupling an initial bootstrapping of user profiles via form filling with a subsequent refinement based on an analysis of user behavior. 
\end{itemize}
Another limitation is related to the questionnaire for gathering user's sensory aversion: it has been derived from a state-of-art SPQ questionnaire, but a correlation between the two has to be computed. We plan to carry out this analysis as well.


\begin{acks}
This work is supported by the COMPAGNIA di SAN PAOLO Foundation. We thank Stefano Cocomazzi, Stefania Brighenti and Claudio Mattutino for their contributions to the work. 
\end{acks}

\bibliographystyle{ACM-Reference-Format}

\begin{thebibliography}{57}


\ifx \showCODEN    \undefined \def \showCODEN     #1{\unskip}     \fi
\ifx \showDOI      \undefined \def \showDOI       #1{#1}\fi
\ifx \showISBNx    \undefined \def \showISBNx     #1{\unskip}     \fi
\ifx \showISBNxiii \undefined \def \showISBNxiii  #1{\unskip}     \fi
\ifx \showISSN     \undefined \def \showISSN      #1{\unskip}     \fi
\ifx \showLCCN     \undefined \def \showLCCN      #1{\unskip}     \fi
\ifx \shownote     \undefined \def \shownote      #1{#1}          \fi
\ifx \showarticletitle \undefined \def \showarticletitle #1{#1}   \fi
\ifx \showURL      \undefined \def \showURL       {\relax}        \fi
\providecommand\bibfield[2]{#2}
\providecommand\bibinfo[2]{#2}
\providecommand\natexlab[1]{#1}
\providecommand\showeprint[2][]{arXiv:#2}

\bibitem[\protect\citeauthoryear{Adomavicius and Kwon}{Adomavicius and
  Kwon}{2007}]%
        {Adomavicius-Kwon:07}
\bibfield{author}{\bibinfo{person}{Gediminas Adomavicius} {and}
  \bibinfo{person}{YoungOk Kwon}.} \bibinfo{year}{2007}\natexlab{}.
\newblock \showarticletitle{New recommendation techniques for multicriteria
  rating systems}.
\newblock \bibinfo{journal}{\emph{IEEE Intelligent Systems}}
  \bibinfo{volume}{22}, \bibinfo{number}{3} (\bibinfo{date}{May}
  \bibinfo{year}{2007}), \bibinfo{pages}{48--55}.
\newblock
\showISSN{1941-1294}
\urldef\tempurl%
\url{https://doi.org/10.1109/MIS.2007.58}
\showDOI{\tempurl}


\bibitem[\protect\citeauthoryear{Adomavicius, Manouselis, and Kwon}{Adomavicius
  et~al\mbox{.}}{2011}]%
        {Adomavicius-Kwon:11}
\bibfield{author}{\bibinfo{person}{Gediminas Adomavicius},
  \bibinfo{person}{Nikos Manouselis}, {and} \bibinfo{person}{YoungOk Kwon}.}
  \bibinfo{year}{2011}\natexlab{}.
\newblock \bibinfo{booktitle}{\emph{Multi-Criteria Recommender Systems}}.
\newblock \bibinfo{publisher}{Springer US}, \bibinfo{address}{Boston, MA},
  \bibinfo{pages}{769--803}.
\newblock
\showISBNx{978-0-387-85820-3}
\urldef\tempurl%
\url{https://doi.org/10.1007/978-0-387-85820-3_24}
\showDOI{\tempurl}


\bibitem[\protect\citeauthoryear{Ardissono, Goy, Petrone, Segnan, and
  Torasso}{Ardissono et~al\mbox{.}}{2003}]%
        {Ardissono-etal:03}
\bibfield{author}{\bibinfo{person}{Liliana Ardissono}, \bibinfo{person}{Anna
  Goy}, \bibinfo{person}{Giovanna Petrone}, \bibinfo{person}{Marino Segnan},
  {and} \bibinfo{person}{Pietro Torasso}.} \bibinfo{year}{2003}\natexlab{}.
\newblock \showarticletitle{{INTRIGUE}: personalized recommendation of tourist
  attractions for desktop and handset devices}.
\newblock \bibinfo{journal}{\emph{Applied Artificial Intelligence, Special
  Issue on Artificial Intelligence for Cultural Heritage and Digital
  Libraries}} \bibinfo{volume}{17}, \bibinfo{number}{8-9}
  (\bibinfo{year}{2003}), \bibinfo{pages}{687--714}.
\newblock
\showISSN{08839514}
\urldef\tempurl%
\url{https://doi.org/10.1080/713827254}
\showDOI{\tempurl}


\bibitem[\protect\citeauthoryear{Association et~al\mbox{.}}{Association
  et~al\mbox{.}}{2013}]%
        {american2013diagnostic}
\bibfield{author}{\bibinfo{person}{American~Psychiatric Association}
  {et~al\mbox{.}}} \bibinfo{year}{2013}\natexlab{}.
\newblock \bibinfo{booktitle}{\emph{Diagnostic and Statistical Manual of Mental
  Disorders (DSM-5{\textregistered})}}.
\newblock \bibinfo{publisher}{American Psychiatric Pub}.
\newblock


\bibitem[\protect\citeauthoryear{Bara, Bucciarelli, and Colle}{Bara
  et~al\mbox{.}}{2001}]%
        {bara2001}
\bibfield{author}{\bibinfo{person}{Bruno~G. Bara}, \bibinfo{person}{Monica
  Bucciarelli}, {and} \bibinfo{person}{Livia Colle}.}
  \bibinfo{year}{2001}\natexlab{}.
\newblock \showarticletitle{Communicative abilities in autism: Evidence for
  attentional deficits}.
\newblock \bibinfo{journal}{\emph{Brain and Language}} \bibinfo{volume}{77},
  \bibinfo{number}{2} (\bibinfo{year}{2001}), \bibinfo{pages}{216 -- 240}.
\newblock
\showISSN{0093-934X}
\urldef\tempurl%
\url{https://doi.org/10.1006/brln.2000.2429}
\showDOI{\tempurl}


\bibitem[\protect\citeauthoryear{Bilici and Saygın}{Bilici and
  Saygın}{2017}]%
        {Bilici-Saygin:17}
\bibfield{author}{\bibinfo{person}{Eda Bilici} {and} \bibinfo{person}{Yücel
  Saygın}.} \bibinfo{year}{2017}\natexlab{}.
\newblock \showarticletitle{Why do people (not) like me?: Mining opinion
  influencing factors from reviews}.
\newblock \bibinfo{journal}{\emph{Expert Systems with Applications}}
  \bibinfo{volume}{68} (\bibinfo{year}{2017}), \bibinfo{pages}{185 -- 195}.
\newblock
\showISSN{0957-4174}
\urldef\tempurl%
\url{https://doi.org/10.1016/j.eswa.2016.10.001}
\showDOI{\tempurl}


\bibitem[\protect\citeauthoryear{Boyd, Rangel, Tomimbang, Conejo-Toledo, Patel,
  Tentori, and Hayes}{Boyd et~al\mbox{.}}{2016}]%
        {boyd2016saywat}
\bibfield{author}{\bibinfo{person}{LouAnne~E. Boyd}, \bibinfo{person}{Alejandro
  Rangel}, \bibinfo{person}{Helen Tomimbang}, \bibinfo{person}{Andrea
  Conejo-Toledo}, \bibinfo{person}{Kanika Patel}, \bibinfo{person}{Monica
  Tentori}, {and} \bibinfo{person}{Gillian~R. Hayes}.}
  \bibinfo{year}{2016}\natexlab{}.
\newblock \showarticletitle{Say{WAT}: Augmenting face-to-face conversations for
  adults with autism}. In \bibinfo{booktitle}{\emph{Proceedings of the 2016 CHI
  Conference on Human Factors in Computing Systems}}
  \emph{(\bibinfo{series}{CHI ’16})}. \bibinfo{publisher}{Association for
  Computing Machinery}, \bibinfo{address}{New York, NY, USA},
  \bibinfo{pages}{4872–4883}.
\newblock
\showISBNx{9781450333627}
\urldef\tempurl%
\url{https://doi.org/10.1145/2858036.2858215}
\showDOI{\tempurl}


\bibitem[\protect\citeauthoryear{Bridge, G\"{o}ker, McGinty, and Smyth}{Bridge
  et~al\mbox{.}}{2005}]%
        {Bridge-etal:05}
\bibfield{author}{\bibinfo{person}{Derek Bridge}, \bibinfo{person}{Mehmet~H.
  G\"{o}ker}, \bibinfo{person}{Lorraine McGinty}, {and} \bibinfo{person}{Barry
  Smyth}.} \bibinfo{year}{2005}\natexlab{}.
\newblock \showarticletitle{Case-based recommender systems}.
\newblock \bibinfo{journal}{\emph{Knowl. Eng. Rev.}} \bibinfo{volume}{20},
  \bibinfo{number}{3} (\bibinfo{date}{Sept.} \bibinfo{year}{2005}),
  \bibinfo{pages}{315--320}.
\newblock
\showISSN{0269-8889}
\urldef\tempurl%
\url{https://doi.org/10.1017/S0269888906000567}
\showDOI{\tempurl}


\bibitem[\protect\citeauthoryear{Burke}{Burke}{2002}]%
        {Burke:02}
\bibfield{author}{\bibinfo{person}{Robin Burke}.}
  \bibinfo{year}{2002}\natexlab{}.
\newblock \showarticletitle{Hybrid recommender systems: survey and
  experiments}.
\newblock \bibinfo{journal}{\emph{User Modeling and User-Adapted Interaction}}
  \bibinfo{volume}{12}, \bibinfo{number}{4} (\bibinfo{year}{2002}),
  \bibinfo{pages}{331--370}.
\newblock
\showISSN{1573-1391}
\urldef\tempurl%
\url{https://doi.org/10.1023/A:1021240730564}
\showDOI{\tempurl}


\bibitem[\protect\citeauthoryear{Cantador, Castells, and
  Bellog\'{\i}n}{Cantador et~al\mbox{.}}{2011}]%
        {Cantador-etal:11}
\bibfield{author}{\bibinfo{person}{Iv{\'a}n Cantador}, \bibinfo{person}{Pablo
  Castells}, {and} \bibinfo{person}{Alejandro Bellog\'{\i}n}.}
  \bibinfo{year}{2011}\natexlab{}.
\newblock \showarticletitle{An enhanced semantic layer for hybrid recommender
  systems: Application to news recommendation}.
\newblock \bibinfo{journal}{\emph{Int. Journal on Semantic Web and Information
  Systems}} \bibinfo{volume}{7}, \bibinfo{number}{1} (\bibinfo{year}{2011}),
  \bibinfo{pages}{44--77}.
\newblock
\showISSN{1552-6283}
\urldef\tempurl%
\url{https://doi.org/10.4018/jswis.2011010103}
\showDOI{\tempurl}


\bibitem[\protect\citeauthoryear{Costa, Costa, Juli{\'a}n, and Novais}{Costa
  et~al\mbox{.}}{2017}]%
        {costa2017task}
\bibfield{author}{\bibinfo{person}{Margarida Costa}, \bibinfo{person}{Angelo
  Costa}, \bibinfo{person}{Vicente Juli{\'a}n}, {and} \bibinfo{person}{Paulo
  Novais}.} \bibinfo{year}{2017}\natexlab{}.
\newblock \showarticletitle{A task recommendation system for children and youth
  with autism spectrum disorder}. In \bibinfo{booktitle}{\emph{Ambient
  Intelligence-- Software and Applications -- 8th International Symposium on
  Ambient Intelligence (ISAmI 2017)}},
  \bibfield{editor}{\bibinfo{person}{Juan~F. De~Paz}, \bibinfo{person}{Vicente
  Juli{\'a}n}, \bibinfo{person}{Gabriel Villarrubia}, \bibinfo{person}{Goreti
  Marreiros}, {and} \bibinfo{person}{Paulo Novais}} (Eds.).
  \bibinfo{publisher}{Springer International Publishing},
  \bibinfo{address}{Cham}, \bibinfo{pages}{87--94}.
\newblock
\showISBNx{978-3-319-61118-1}
\urldef\tempurl%
\url{https://doi.org/10.1007/978-3-319-61118-1_12}
\showDOI{\tempurl}


\bibitem[\protect\citeauthoryear{Dragone, Pellegrini, Vescovi, Tentori, and
  Passerini}{Dragone et~al\mbox{.}}{2018}]%
        {Dragone-etal:18}
\bibfield{author}{\bibinfo{person}{Paolo Dragone}, \bibinfo{person}{Giovanni
  Pellegrini}, \bibinfo{person}{Michele Vescovi}, \bibinfo{person}{Katya
  Tentori}, {and} \bibinfo{person}{Andrea Passerini}.}
  \bibinfo{year}{2018}\natexlab{}.
\newblock \showarticletitle{No more ready-made deals: constructive
  recommendation for telco service bundling}. In
  \bibinfo{booktitle}{\emph{Proceedings of the 12th ACM Conference on
  Recommender Systems}} \emph{(\bibinfo{series}{RecSys '18})}.
  \bibinfo{publisher}{ACM}, \bibinfo{address}{New York, NY, USA},
  \bibinfo{pages}{163--171}.
\newblock
\showISBNx{978-1-4503-5901-6}
\urldef\tempurl%
\url{https://doi.org/10.1145/3240323.3240348}
\showDOI{\tempurl}


\bibitem[\protect\citeauthoryear{Elsabbagh, Divan, Koh, Kim, Kauchali,
  Marc{\'\i}n, Montiel-Nava, Patel, Paula, Wang, et~al\mbox{.}}{Elsabbagh
  et~al\mbox{.}}{2012}]%
        {elsabbagh2012global}
\bibfield{author}{\bibinfo{person}{Mayada Elsabbagh}, \bibinfo{person}{Gauri
  Divan}, \bibinfo{person}{Yun-Joo Koh}, \bibinfo{person}{Young~Shin Kim},
  \bibinfo{person}{Shuaib Kauchali}, \bibinfo{person}{Carlos Marc{\'\i}n},
  \bibinfo{person}{Cecilia Montiel-Nava}, \bibinfo{person}{Vikram Patel},
  \bibinfo{person}{Cristiane~S. Paula}, \bibinfo{person}{Chongying Wang},
  {et~al\mbox{.}}} \bibinfo{year}{2012}\natexlab{}.
\newblock \showarticletitle{Global prevalence of autism and other pervasive
  developmental disorders}.
\newblock \bibinfo{journal}{\emph{Autism research}} \bibinfo{volume}{5},
  \bibinfo{number}{3} (\bibinfo{year}{2012}), \bibinfo{pages}{160--179}.
\newblock
\urldef\tempurl%
\url{https://doi.org/10.1002/aur.239}
\showDOI{\tempurl}


\bibitem[\protect\citeauthoryear{Felfernig, Friedrich, Jannach, and
  Zanker}{Felfernig et~al\mbox{.}}{2011}]%
        {Felfernig-etal:11}
\bibfield{author}{\bibinfo{person}{Alexander Felfernig},
  \bibinfo{person}{Gerhard Friedrich}, \bibinfo{person}{Dietmar Jannach}, {and}
  \bibinfo{person}{Markus Zanker}.} \bibinfo{year}{2011}\natexlab{}.
\newblock \bibinfo{booktitle}{\emph{Developing Constraint-based Recommenders}}.
\newblock \bibinfo{publisher}{Springer US}, \bibinfo{address}{Boston, MA},
  \bibinfo{pages}{187--215}.
\newblock
\showISBNx{978-0-387-85820-3}
\urldef\tempurl%
\url{https://doi.org/10.1007/978-0-387-85820-3_6}
\showDOI{\tempurl}


\bibitem[\protect\citeauthoryear{Freyne and Berkovsky}{Freyne and
  Berkovsky}{2010}]%
        {Freyne-Berkovsky:10}
\bibfield{author}{\bibinfo{person}{Jill Freyne} {and} \bibinfo{person}{Shlomo
  Berkovsky}.} \bibinfo{year}{2010}\natexlab{}.
\newblock \showarticletitle{Intelligent food planning: Personalized recipe
  recommendation}. In \bibinfo{booktitle}{\emph{Proceedings of the 15th
  International Conference on Intelligent User Interfaces}}
  \emph{(\bibinfo{series}{IUI '10})}. \bibinfo{publisher}{ACM},
  \bibinfo{address}{New York, NY, USA}, \bibinfo{pages}{321--324}.
\newblock
\showISBNx{978-1-60558-515-4}
\urldef\tempurl%
\url{https://doi.org/10.1145/1719970.1720021}
\showDOI{\tempurl}


\bibitem[\protect\citeauthoryear{Garc{\'{\i}}a, Stash, Fabri, Bra, Fletcher,
  and Pechenizkiy}{Garc{\'{\i}}a et~al\mbox{.}}{2016}]%
        {DBLP:conf/ht/GarciaSFBFP16}
\bibfield{author}{\bibinfo{person}{Alejandro~Montes Garc{\'{\i}}a},
  \bibinfo{person}{Natalia Stash}, \bibinfo{person}{Marc Fabri},
  \bibinfo{person}{Paul~De Bra}, \bibinfo{person}{George H.~L. Fletcher}, {and}
  \bibinfo{person}{Mykola Pechenizkiy}.} \bibinfo{year}{2016}\natexlab{}.
\newblock \showarticletitle{Adaptive web-based educational application for
  autistic students}. In \bibinfo{booktitle}{\emph{Late-breaking Results,
  Demos, Doctoral Consortium, Workshops Proceedings and Creative Track of the
  27th {ACM} Conference on Hypertext and Social Media {(HT} 2016), Halifax,
  Canada, July 13-16, 2016}} \emph{(\bibinfo{series}{{CEUR} Workshop
  Proceedings})}, \bibfield{editor}{\bibinfo{person}{Kevin Koidl} {and}
  \bibinfo{person}{Ben Steichen}} (Eds.), Vol.~\bibinfo{volume}{1628}.
  \bibinfo{publisher}{CEUR-WS.org}.
\newblock
\urldef\tempurl%
\url{http://ceur-ws.org/Vol-1628/Demo1.pdf}
\showURL{%
\tempurl}


\bibitem[\protect\citeauthoryear{Gemmell, Schimoler, Mobasher, and
  Burke}{Gemmell et~al\mbox{.}}{2012}]%
        {Gemmel-etal:12}
\bibfield{author}{\bibinfo{person}{Jonathan Gemmell}, \bibinfo{person}{Thomas
  Schimoler}, \bibinfo{person}{Bamshad Mobasher}, {and} \bibinfo{person}{Robin
  Burke}.} \bibinfo{year}{2012}\natexlab{}.
\newblock \showarticletitle{Resource recommendation in social annotation
  systems: A linear-weighted hybrid approach}.
\newblock \bibinfo{journal}{\emph{J. Comput. System Sci.}}
  \bibinfo{volume}{78}, \bibinfo{number}{4} (\bibinfo{year}{2012}),
  \bibinfo{pages}{1160 -- 1174}.
\newblock
\showISSN{0022-0000}
\urldef\tempurl%
\url{https://doi.org/10.1016/j.jcss.2011.10.006}
\showDOI{\tempurl}


\bibitem[\protect\citeauthoryear{Gillott and Standen}{Gillott and
  Standen}{2007}]%
        {gillott2007levels}
\bibfield{author}{\bibinfo{person}{Alinda Gillott} {and} \bibinfo{person}{PJ
  Standen}.} \bibinfo{year}{2007}\natexlab{}.
\newblock \showarticletitle{Levels of anxiety and sources of stress in adults
  with autism}.
\newblock \bibinfo{journal}{\emph{Journal of Intellectual Disabilities}}
  \bibinfo{volume}{11}, \bibinfo{number}{4} (\bibinfo{year}{2007}),
  \bibinfo{pages}{359--370}.
\newblock
\urldef\tempurl%
\url{https://doi.org/10.1177/1744629507083585}
\showDOI{\tempurl}


\bibitem[\protect\citeauthoryear{Golan, Ashwin, Granader, McClintock, Day,
  Leggett, and Baron-Cohen}{Golan et~al\mbox{.}}{2010}]%
        {golan2010enhancing}
\bibfield{author}{\bibinfo{person}{Ofer Golan}, \bibinfo{person}{Emma Ashwin},
  \bibinfo{person}{Yael Granader}, \bibinfo{person}{Suzy McClintock},
  \bibinfo{person}{Kate Day}, \bibinfo{person}{Victoria Leggett}, {and}
  \bibinfo{person}{Simon Baron-Cohen}.} \bibinfo{year}{2010}\natexlab{}.
\newblock \showarticletitle{Enhancing emotion recognition in children with
  autism spectrum conditions: An intervention using animated vehicles with real
  emotional faces}.
\newblock \bibinfo{journal}{\emph{Journal of Autism and Developmental
  Disorders}} \bibinfo{volume}{40}, \bibinfo{number}{3} (\bibinfo{year}{2010}),
  \bibinfo{pages}{269--279}.
\newblock
\urldef\tempurl%
\url{https://doi.org/10.1007/s10803-009-0862-9}
\showDOI{\tempurl}


\bibitem[\protect\citeauthoryear{Goldsmith and LeBlanc}{Goldsmith and
  LeBlanc}{2004}]%
        {goldsmith2004use}
\bibfield{author}{\bibinfo{person}{Tina~R. Goldsmith} {and}
  \bibinfo{person}{Linda~A. LeBlanc}.} \bibinfo{year}{2004}\natexlab{}.
\newblock \showarticletitle{Use of technology in interventions for children
  with autism}.
\newblock \bibinfo{journal}{\emph{Journal of Early and Intensive Behavior
  Intervention}} \bibinfo{volume}{1}, \bibinfo{number}{2}
  (\bibinfo{year}{2004}), \bibinfo{pages}{166}.
\newblock
\urldef\tempurl%
\url{https://doi.org/10.1037/h0100287}
\showDOI{\tempurl}


\bibitem[\protect\citeauthoryear{Grynszpan, Weiss, Perez-Diaz, and
  Gal}{Grynszpan et~al\mbox{.}}{2014}]%
        {grynszpan2014innovative}
\bibfield{author}{\bibinfo{person}{Ouriel Grynszpan},
  \bibinfo{person}{Patrice~L Weiss}, \bibinfo{person}{Fernando Perez-Diaz},
  {and} \bibinfo{person}{Eynat Gal}.} \bibinfo{year}{2014}\natexlab{}.
\newblock \showarticletitle{Innovative technology-based interventions for
  autism spectrum disorders: a meta-analysis}.
\newblock \bibinfo{journal}{\emph{Autism}} \bibinfo{volume}{18},
  \bibinfo{number}{4} (\bibinfo{year}{2014}), \bibinfo{pages}{346--361}.
\newblock
\urldef\tempurl%
\url{https://doi.org/10.1177/1362361313476767}
\showDOI{\tempurl}


\bibitem[\protect\citeauthoryear{Han and Karypis}{Han and Karypis}{2005}]%
        {Han-etal:05}
\bibfield{author}{\bibinfo{person}{Eui-Hong~(Sam) Han} {and}
  \bibinfo{person}{George Karypis}.} \bibinfo{year}{2005}\natexlab{}.
\newblock \showarticletitle{Feature-based recommendation system}. In
  \bibinfo{booktitle}{\emph{Proceedings of the 14th ACM International
  Conference on Information and Knowledge Management}}
  \emph{(\bibinfo{series}{CIKM '05})}. \bibinfo{publisher}{ACM},
  \bibinfo{address}{New York, NY, USA}, \bibinfo{pages}{446--452}.
\newblock


\bibitem[\protect\citeauthoryear{Hobson}{Hobson}{1995}]%
        {hobson1995}
\bibfield{author}{\bibinfo{person}{R.~Peter Hobson}.}
  \bibinfo{year}{1995}\natexlab{}.
\newblock \bibinfo{booktitle}{\emph{Autism and the development of mind}}.
\newblock \bibinfo{publisher}{Routledge}.
\newblock


\bibitem[\protect\citeauthoryear{Hong, Kim, Abowd, and Arriaga}{Hong
  et~al\mbox{.}}{2012}]%
        {hong2012designing}
\bibfield{author}{\bibinfo{person}{Hwajung Hong}, \bibinfo{person}{Jennifer~G.
  Kim}, \bibinfo{person}{Gregory~D. Abowd}, {and} \bibinfo{person}{Rosa~I.
  Arriaga}.} \bibinfo{year}{2012}\natexlab{}.
\newblock \showarticletitle{Designing a social network to support the
  independence of young adults with autism}. In
  \bibinfo{booktitle}{\emph{Proceedings of the ACM 2012 Conference on Computer
  Supported Cooperative Work}} \emph{(\bibinfo{series}{CSCW ’12})}.
  \bibinfo{publisher}{Association for Computing Machinery},
  \bibinfo{address}{New York, NY, USA}, \bibinfo{pages}{627–636}.
\newblock
\showISBNx{9781450310864}
\urldef\tempurl%
\url{https://doi.org/10.1145/2145204.2145300}
\showDOI{\tempurl}


\bibitem[\protect\citeauthoryear{Jannach, Zanker, and Fuchs}{Jannach
  et~al\mbox{.}}{2014}]%
        {Jannach-etal:14}
\bibfield{author}{\bibinfo{person}{Dietmar Jannach}, \bibinfo{person}{Markus
  Zanker}, {and} \bibinfo{person}{Matthias Fuchs}.}
  \bibinfo{year}{2014}\natexlab{}.
\newblock \showarticletitle{Leveraging multi-criteria customer feedback for
  satisfaction analysis and improved recommendations}.
\newblock \bibinfo{journal}{\emph{Information Technology {\&} Tourism}}
  \bibinfo{volume}{14}, \bibinfo{number}{2} (\bibinfo{date}{01 Jul}
  \bibinfo{year}{2014}), \bibinfo{pages}{119--149}.
\newblock
\showISSN{1943-4294}
\urldef\tempurl%
\url{https://doi.org/10.1007/s40558-014-0010-z}
\showDOI{\tempurl}


\bibitem[\protect\citeauthoryear{Judy, Krishnakumar, and Narayanan}{Judy
  et~al\mbox{.}}{2012}]%
        {judy2012}
\bibfield{author}{\bibinfo{person}{M.V. Judy}, \bibinfo{person}{U.
  Krishnakumar}, {and} \bibinfo{person}{A.G.~Hari Narayanan}.}
  \bibinfo{year}{2012}\natexlab{}.
\newblock \showarticletitle{Constructing a personalized e-learning system for
  students with autism based on soft semantic web technologies}. In
  \bibinfo{booktitle}{\emph{2012 IEEE International Conference on Technology
  Enhanced Education (ICTEE)}}. IEEE, \bibinfo{pages}{1--5}.
\newblock
\urldef\tempurl%
\url{https://doi.org/10.1109/ICTEE.2012.6208625}
\showDOI{\tempurl}


\bibitem[\protect\citeauthoryear{Kientz, Goodwin, Hayes, and Abowd}{Kientz
  et~al\mbox{.}}{2013}]%
        {kientz2013interactive}
\bibfield{author}{\bibinfo{person}{Julie~A Kientz}, \bibinfo{person}{Matthew~S
  Goodwin}, \bibinfo{person}{Gillian~R Hayes}, {and} \bibinfo{person}{Gregory~D
  Abowd}.} \bibinfo{year}{2013}\natexlab{}.
\newblock \showarticletitle{Interactive technologies for autism}.
\newblock \bibinfo{journal}{\emph{Synthesis Lectures on Assistive,
  Rehabilitative, and Health-Preserving Technologies}} \bibinfo{volume}{2},
  \bibinfo{number}{2} (\bibinfo{year}{2013}), \bibinfo{pages}{1--177}.
\newblock
\urldef\tempurl%
\url{https://doi.org/10.2200/S00533ED1V01Y201309ARH004}
\showDOI{\tempurl}


\bibitem[\protect\citeauthoryear{Lops, {de Gemmis}, and Semeraro}{Lops
  et~al\mbox{.}}{2011}]%
        {Lops-etal:11}
\bibfield{author}{\bibinfo{person}{Pasquale Lops}, \bibinfo{person}{Marco {de
  Gemmis}}, {and} \bibinfo{person}{Giovanni Semeraro}.}
  \bibinfo{year}{2011}\natexlab{}.
\newblock \bibinfo{booktitle}{\emph{Content-based Recommender Systems: State of
  the Art and Trends}}.
\newblock \bibinfo{publisher}{Springer US}, \bibinfo{address}{Boston, MA},
  \bibinfo{pages}{73--105}.
\newblock
\showISBNx{978-0-387-85820-3}
\urldef\tempurl%
\url{https://doi.org/10.1007/978-0-387-85820-3_3}
\showDOI{\tempurl}


\bibitem[\protect\citeauthoryear{Murray, Lesser, and Lawson}{Murray
  et~al\mbox{.}}{2005}]%
        {murray2005attention}
\bibfield{author}{\bibinfo{person}{Dinah Murray}, \bibinfo{person}{Mike
  Lesser}, {and} \bibinfo{person}{Wendy Lawson}.}
  \bibinfo{year}{2005}\natexlab{}.
\newblock \showarticletitle{Attention, monotropism and the diagnostic criteria
  for autism}.
\newblock \bibinfo{journal}{\emph{Autism}} \bibinfo{volume}{9},
  \bibinfo{number}{2} (\bibinfo{year}{2005}), \bibinfo{pages}{139--156}.
\newblock
\urldef\tempurl%
\url{https://doi.org/10.1177/1362361305051398}
\showDOI{\tempurl}


\bibitem[\protect\citeauthoryear{Musto, Semeraro, Lops, and de~Gemmis}{Musto
  et~al\mbox{.}}{2011}]%
        {Musto-etal:11}
\bibfield{author}{\bibinfo{person}{Cataldo Musto}, \bibinfo{person}{Giovanni
  Semeraro}, \bibinfo{person}{Pasquale Lops}, {and} \bibinfo{person}{Marco de
  Gemmis}.} \bibinfo{year}{2011}\natexlab{}.
\newblock \showarticletitle{Random indexing and negative user preferences for
  enhancing content-based recommender systems}. In
  \bibinfo{booktitle}{\emph{E-Commerce and Web Technologies}},
  \bibfield{editor}{\bibinfo{person}{Christian Huemer} {and}
  \bibinfo{person}{Thomas Setzer}} (Eds.). \bibinfo{publisher}{Springer Berlin
  Heidelberg}, \bibinfo{address}{Berlin, Heidelberg},
  \bibinfo{pages}{270--281}.
\newblock
\showISBNx{978-3-642-23013-4}
\urldef\tempurl%
\url{https://doi.org/10.1007/978-3-642-23014-1_23}
\showDOI{\tempurl}


\bibitem[\protect\citeauthoryear{Ng and Pera}{Ng and Pera}{2018}]%
        {Ng-Pera:18}
\bibfield{author}{\bibinfo{person}{{Yiu-Kai} Ng} {and} \bibinfo{person}{{Maria
  Soledad} Pera}.} \bibinfo{year}{2018}\natexlab{}.
\newblock \showarticletitle{Recommending social-interactive games for adults
  with autism spectrum disorders {(ASD)}}. In
  \bibinfo{booktitle}{\emph{Proceedings of the 12th ACM Conference on
  Recommender Systems}} \emph{(\bibinfo{series}{RecSys '18})}.
  \bibinfo{publisher}{ACM}, \bibinfo{address}{New York, NY, USA},
  \bibinfo{pages}{209--213}.
\newblock
\showISBNx{978-1-4503-5901-6}
\urldef\tempurl%
\url{https://doi.org/10.1145/3240323.3240405}
\showDOI{\tempurl}


\bibitem[\protect\citeauthoryear{{OpenStreetMap Contributors}}{{OpenStreetMap
  Contributors}}{2017}]%
        {OpenStreetMap}
\bibfield{author}{\bibinfo{person}{{OpenStreetMap Contributors}}.}
  \bibinfo{year}{2017}\natexlab{}.
\newblock \showarticletitle{Openstreetmap}.
  \bibinfo{address}{\url{https://www.openstreetmap.org}}.
\newblock


\bibitem[\protect\citeauthoryear{Pera and Ng}{Pera and Ng}{2014}]%
        {pera2014automating}
\bibfield{author}{\bibinfo{person}{Maria~Soledad Pera} {and}
  \bibinfo{person}{Yiu-Kai Ng}.} \bibinfo{year}{2014}\natexlab{}.
\newblock \showarticletitle{Automating readers’ advisory to make book
  recommendations for k-12 readers}. In \bibinfo{booktitle}{\emph{Proceedings
  of the 8th ACM Conference on Recommender Systems}}
  \emph{(\bibinfo{series}{RecSys ’14})}. \bibinfo{publisher}{Association for
  Computing Machinery}, \bibinfo{address}{New York, NY, USA},
  \bibinfo{pages}{9–16}.
\newblock
\showISBNx{9781450326681}
\urldef\tempurl%
\url{https://doi.org/10.1145/2645710.2645721}
\showDOI{\tempurl}


\bibitem[\protect\citeauthoryear{Premasundari and Yamini}{Premasundari and
  Yamini}{2019}]%
        {premasundari2019food}
\bibfield{author}{\bibinfo{person}{M. Premasundari} {and} \bibinfo{person}{C.
  Yamini}.} \bibinfo{year}{2019}\natexlab{}.
\newblock \showarticletitle{Food and therapy recommendation system for autistic
  syndrome using machine learning techniques}. In
  \bibinfo{booktitle}{\emph{2019 IEEE International Conference on Electrical,
  Computer and Communication Technologies (ICECCT)}}. IEEE,
  \bibinfo{pages}{1--6}.
\newblock
\urldef\tempurl%
\url{https://doi.org/10.1109/ICECCT.2019.8868979}
\showDOI{\tempurl}


\bibitem[\protect\citeauthoryear{Putnam and Chong}{Putnam and Chong}{2008}]%
        {putnam2008software}
\bibfield{author}{\bibinfo{person}{Cynthia Putnam} {and} \bibinfo{person}{Lorna
  Chong}.} \bibinfo{year}{2008}\natexlab{}.
\newblock \showarticletitle{Software and technologies designed for people with
  autism: What do users want?}. In \bibinfo{booktitle}{\emph{Proceedings of the
  10th International ACM SIGACCESS Conference on Computers and Accessibility}}
  \emph{(\bibinfo{series}{Assets ’08})}. \bibinfo{publisher}{Association for
  Computing Machinery}, \bibinfo{address}{New York, NY, USA},
  \bibinfo{pages}{3–10}.
\newblock
\showISBNx{9781595939760}
\urldef\tempurl%
\url{https://doi.org/10.1145/1414471.1414475}
\showDOI{\tempurl}


\bibitem[\protect\citeauthoryear{Putnam, Hanschke, Todd, Gemmell, and
  Kollia}{Putnam et~al\mbox{.}}{2019}]%
        {putnam2019interactive}
\bibfield{author}{\bibinfo{person}{Cynthia Putnam}, \bibinfo{person}{Christina
  Hanschke}, \bibinfo{person}{Jennifer Todd}, \bibinfo{person}{Jonathan
  Gemmell}, {and} \bibinfo{person}{Mia Kollia}.}
  \bibinfo{year}{2019}\natexlab{}.
\newblock \showarticletitle{Interactive technologies designed for children with
  autism: Reports of use and desires from parents, teachers, and therapists}.
\newblock \bibinfo{journal}{\emph{ACM Trans. Access. Comput.}}
  \bibinfo{volume}{12}, \bibinfo{number}{3}, Article \bibinfo{articleno}{12}
  (\bibinfo{date}{Sept.} \bibinfo{year}{2019}), \bibinfo{numpages}{37}~pages.
\newblock
\showISSN{1936-7228}
\urldef\tempurl%
\url{https://doi.org/10.1145/3342285}
\showDOI{\tempurl}


\bibitem[\protect\citeauthoryear{Quercia, Schifanella, and Aiello}{Quercia
  et~al\mbox{.}}{2014}]%
        {Quercia-etal:14}
\bibfield{author}{\bibinfo{person}{Daniele Quercia}, \bibinfo{person}{Rossano
  Schifanella}, {and} \bibinfo{person}{Luca~Maria Aiello}.}
  \bibinfo{year}{2014}\natexlab{}.
\newblock \showarticletitle{The shortest path to happiness: recommending
  beautiful, quiet, and happy routes in the city}. In
  \bibinfo{booktitle}{\emph{Proceedings of the 25th ACM Conference on Hypertext
  and Social Media}} \emph{(\bibinfo{series}{HT '14})}.
  \bibinfo{publisher}{ACM}, \bibinfo{address}{New York, NY, USA},
  \bibinfo{pages}{116--125}.
\newblock
\showISBNx{978-1-4503-2954-5}
\urldef\tempurl%
\url{https://doi.org/10.1145/2631775.2631799}
\showDOI{\tempurl}


\bibitem[\protect\citeauthoryear{Ramdoss, Machalicek, Rispoli, Mulloy, Lang,
  and O’Reilly}{Ramdoss et~al\mbox{.}}{2012}]%
        {ramdoss2012computer}
\bibfield{author}{\bibinfo{person}{Sathiyaprakash Ramdoss},
  \bibinfo{person}{Wendy Machalicek}, \bibinfo{person}{Mandy Rispoli},
  \bibinfo{person}{Austin Mulloy}, \bibinfo{person}{Russell Lang}, {and}
  \bibinfo{person}{Mark O’Reilly}.} \bibinfo{year}{2012}\natexlab{}.
\newblock \showarticletitle{Computer-based interventions to improve social and
  emotional skills in individuals with autism spectrum disorders: a systematic
  review}.
\newblock \bibinfo{journal}{\emph{Developmental Neurorehabilitation}}
  \bibinfo{volume}{15}, \bibinfo{number}{2} (\bibinfo{year}{2012}),
  \bibinfo{pages}{119--135}.
\newblock
\urldef\tempurl%
\url{https://doi.org/10.3109/17518423.2011.651655}
\showDOI{\tempurl}
\showeprint{https://doi.org/10.3109/17518423.2011.651655}
\newblock
\shownote{PMID: 22494084.}


\bibitem[\protect\citeauthoryear{Rapp, Cena, Boella, Antonini, Calafiore,
  Buccoliero, Tirassa, Keller, Castaldo, and Brighenti}{Rapp
  et~al\mbox{.}}{2017}]%
        {DBLP:conf/chi/RappCBACBTKCB17}
\bibfield{author}{\bibinfo{person}{Amon Rapp}, \bibinfo{person}{Federica Cena},
  \bibinfo{person}{Guido Boella}, \bibinfo{person}{Alessio Antonini},
  \bibinfo{person}{Alessia Calafiore}, \bibinfo{person}{Stefania Buccoliero},
  \bibinfo{person}{Maurizio Tirassa}, \bibinfo{person}{Roberto Keller},
  \bibinfo{person}{Romina Castaldo}, {and} \bibinfo{person}{Stefania
  Brighenti}.} \bibinfo{year}{2017}\natexlab{}.
\newblock \showarticletitle{Interactive urban maps for people with autism
  spectrum disorder}. In \bibinfo{booktitle}{\emph{Proceedings of the 2017
  {CHI} Conference on Human Factors in Computing Systems, Denver, CO, USA, May
  06-11, 2017, Extended Abstracts}}. \bibinfo{pages}{1987--1992}.
\newblock
\urldef\tempurl%
\url{https://doi.org/10.1145/3027063.3053145}
\showDOI{\tempurl}


\bibitem[\protect\citeauthoryear{Rapp, Cena, Castaldo, Keller, and
  Tirassa}{Rapp et~al\mbox{.}}{2018}]%
        {rapp2018designing}
\bibfield{author}{\bibinfo{person}{Amon Rapp}, \bibinfo{person}{Federica Cena},
  \bibinfo{person}{Romina Castaldo}, \bibinfo{person}{Roberto Keller}, {and}
  \bibinfo{person}{Maurizio Tirassa}.} \bibinfo{year}{2018}\natexlab{}.
\newblock \showarticletitle{Designing technology for spatial needs: Routines,
  control and social competences of people with autism}.
\newblock \bibinfo{journal}{\emph{International Journal of Human-Computer
  Studies}}  \bibinfo{volume}{120} (\bibinfo{year}{2018}), \bibinfo{pages}{49
  -- 65}.
\newblock
\showISSN{1071-5819}
\urldef\tempurl%
\url{https://doi.org/10.1016/j.ijhcs.2018.07.005}
\showDOI{\tempurl}


\bibitem[\protect\citeauthoryear{Rapp, Cena, Mattutino, Boella, Schifanella,
  Keller, and Brighenti}{Rapp et~al\mbox{.}}{2019}]%
        {DBLP:conf/mhci/RappCMBSKB19}
\bibfield{author}{\bibinfo{person}{Amon Rapp}, \bibinfo{person}{Federica Cena},
  \bibinfo{person}{Claudio Mattutino}, \bibinfo{person}{Guido Boella},
  \bibinfo{person}{Claudio Schifanella}, \bibinfo{person}{Roberto Keller},
  {and} \bibinfo{person}{Stefania Brighenti}.} \bibinfo{year}{2019}\natexlab{}.
\newblock \showarticletitle{Designing an urban support for autism}. In
  \bibinfo{booktitle}{\emph{Proceedings of the 21st International Conference on
  Human-Computer Interaction with Mobile Devices and Services, MobileHCI 2019,
  Taipei, Taiwan, October 1-4, 2019}}. \bibinfo{pages}{43:1--43:6}.
\newblock
\urldef\tempurl%
\url{https://doi.org/10.1145/3338286.3344390}
\showDOI{\tempurl}


\bibitem[\protect\citeauthoryear{Rapp, Cena, Schifanella, and Boella}{Rapp
  et~al\mbox{.}}{2020}]%
        {rapp2019spatial}
\bibfield{author}{\bibinfo{person}{Amon Rapp}, \bibinfo{person}{Federica Cena},
  \bibinfo{person}{Claudio Schifanella}, {and} \bibinfo{person}{Guido Boella}.}
  \bibinfo{year}{2020}\natexlab{}.
\newblock \showarticletitle{Finding a secure place: A map-based crowdsourcing
  system for people with autism}.
\newblock \bibinfo{journal}{\emph{IEEE Transactions on Human-Machine Systems}}
  (\bibinfo{year}{2020}).
\newblock


\bibitem[\protect\citeauthoryear{Ricci, Rokach, and Shapira}{Ricci
  et~al\mbox{.}}{2011}]%
        {Ricci-etal:11}
\bibfield{author}{\bibinfo{person}{Francesco Ricci}, \bibinfo{person}{Lior
  Rokach}, {and} \bibinfo{person}{Bracha Shapira}.}
  \bibinfo{year}{2011}\natexlab{}.
\newblock \bibinfo{booktitle}{\emph{Introduction to Recommender Systems
  Handbook}}.
\newblock \bibinfo{publisher}{Springer US}, \bibinfo{address}{Boston, MA},
  \bibinfo{pages}{1--35}.
\newblock
\showISBNx{978-0-387-85820-3}
\urldef\tempurl%
\url{https://doi.org/10.1007/978-0-387-85820-3\_1}
\showDOI{\tempurl}


\bibitem[\protect\citeauthoryear{Robertson and Simmons}{Robertson and
  Simmons}{2013}]%
        {robertson2013relationship}
\bibfield{author}{\bibinfo{person}{Ashley~E Robertson} {and}
  \bibinfo{person}{David~R Simmons}.} \bibinfo{year}{2013}\natexlab{}.
\newblock \showarticletitle{The relationship between sensory sensitivity and
  autistic traits in the general population}.
\newblock \bibinfo{journal}{\emph{Journal of Autism and Developmental
  disorders}} \bibinfo{volume}{43}, \bibinfo{number}{4} (\bibinfo{year}{2013}),
  \bibinfo{pages}{775--784}.
\newblock
\urldef\tempurl%
\url{https://doi.org/10.1007/s10803-012-1608-7}
\showDOI{\tempurl}


\bibitem[\protect\citeauthoryear{Robertson and Baron-Cohen}{Robertson and
  Baron-Cohen}{2017}]%
        {robertson2017sensory}
\bibfield{author}{\bibinfo{person}{Caroline~E. Robertson} {and}
  \bibinfo{person}{Simon Baron-Cohen}.} \bibinfo{year}{2017}\natexlab{}.
\newblock \showarticletitle{Sensory perception in autism}.
\newblock \bibinfo{journal}{\emph{Nature Reviews Neuroscience}}
  \bibinfo{volume}{18}, \bibinfo{number}{11} (\bibinfo{year}{2017}),
  \bibinfo{pages}{671}.
\newblock
\urldef\tempurl%
\url{https://doi.org/10.1038/nrn.2017.112}
\showDOI{\tempurl}


\bibitem[\protect\citeauthoryear{Schopler and Mesibov}{Schopler and
  Mesibov}{1986}]%
        {schopler1986social}
\bibfield{author}{\bibinfo{person}{Eric Schopler} {and}
  \bibinfo{person}{Gary~B. Mesibov}.} \bibinfo{year}{1986}\natexlab{}.
\newblock \bibinfo{booktitle}{\emph{Social Behavior in Autism}}.
\newblock \bibinfo{publisher}{Springer Science \& Business Media}.
\newblock


\bibitem[\protect\citeauthoryear{Simm, Ferrario, Gradinar, Tavares~Smith,
  Forshaw, Smith, and Whittle}{Simm et~al\mbox{.}}{2016}]%
        {simm2016anxiety}
\bibfield{author}{\bibinfo{person}{Will Simm}, \bibinfo{person}{Maria~Angela
  Ferrario}, \bibinfo{person}{Adrian Gradinar}, \bibinfo{person}{Marcia
  Tavares~Smith}, \bibinfo{person}{Stephen Forshaw}, \bibinfo{person}{Ian
  Smith}, {and} \bibinfo{person}{Jon Whittle}.}
  \bibinfo{year}{2016}\natexlab{}.
\newblock \showarticletitle{Anxiety and autism: Towards personalized digital
  health}. In \bibinfo{booktitle}{\emph{Proceedings of the 2016 CHI Conference
  on Human Factors in Computing Systems}} \emph{(\bibinfo{series}{CHI ’16})}.
  \bibinfo{publisher}{Association for Computing Machinery},
  \bibinfo{address}{New York, NY, USA}, \bibinfo{pages}{1270–1281}.
\newblock
\showISBNx{9781450333627}
\urldef\tempurl%
\url{https://doi.org/10.1145/2858036.2858259}
\showDOI{\tempurl}


\bibitem[\protect\citeauthoryear{Smith}{Smith}{2015}]%
        {smith2015spatial}
\bibfield{author}{\bibinfo{person}{Alastair~D. Smith}.}
  \bibinfo{year}{2015}\natexlab{}.
\newblock \showarticletitle{Spatial navigation in autism spectrum disorders: a
  critical review}.
\newblock \bibinfo{journal}{\emph{Frontiers in Psychology}}
  \bibinfo{volume}{6} (\bibinfo{year}{2015}), \bibinfo{pages}{31}.
\newblock
\showISSN{1664-1078}
\urldef\tempurl%
\url{https://doi.org/10.3389/fpsyg.2015.00031}
\showDOI{\tempurl}


\bibitem[\protect\citeauthoryear{{Social Computing}}{{Social
  Computing}}{2020}]%
        {Maps4All}
\bibfield{author}{\bibinfo{person}{{Social Computing}}.}
  \bibinfo{year}{2020}\natexlab{}.
\newblock \showarticletitle{Firstlife - maps4all}.
  \bibinfo{address}{\url{https://maps4all.firstlife.org}}.
\newblock


\bibitem[\protect\citeauthoryear{Tavassoli, Hoekstra, and
  Baron-Cohen}{Tavassoli et~al\mbox{.}}{2014a}]%
        {tavassoli_sensoriality}
\bibfield{author}{\bibinfo{person}{Teresa Tavassoli},
  \bibinfo{person}{Rosa~{A.} Hoekstra}, {and} \bibinfo{person}{Simon
  Baron-Cohen}.} \bibinfo{year}{2014}\natexlab{a}.
\newblock \showarticletitle{The sensory perception quotient (spq): Development
  and validation of a new sensory questionnaire for adults with and without
  autism}.
\newblock \bibinfo{journal}{\emph{Molecular Autism}}  \bibinfo{volume}{5}
  (\bibinfo{year}{2014}), \bibinfo{pages}{29}.
\newblock
\urldef\tempurl%
\url{https://doi.org/10.1186/2040-2392-5-29}
\showDOI{\tempurl}


\bibitem[\protect\citeauthoryear{Tavassoli, Miller, Schoen, Nielsen, and
  Baron-Cohen}{Tavassoli et~al\mbox{.}}{2014b}]%
        {tavassoli2014sensory}
\bibfield{author}{\bibinfo{person}{Teresa Tavassoli}, \bibinfo{person}{Lucy~J.
  Miller}, \bibinfo{person}{Sarah~A. Schoen}, \bibinfo{person}{Darci~M.
  Nielsen}, {and} \bibinfo{person}{Simon Baron-Cohen}.}
  \bibinfo{year}{2014}\natexlab{b}.
\newblock \showarticletitle{Sensory over-responsivity in adults with autism
  spectrum conditions}.
\newblock \bibinfo{journal}{\emph{Autism}} \bibinfo{volume}{18},
  \bibinfo{number}{4} (\bibinfo{year}{2014}), \bibinfo{pages}{428--432}.
\newblock
\urldef\tempurl%
\url{https://doi.org/10.1177/1362361313477246}
\showDOI{\tempurl}
\newblock
\shownote{PMID: 24085741.}


\bibitem[\protect\citeauthoryear{TripAdvisor}{TripAdvisor}{2017}]%
        {TripAdvisor}
\bibfield{author}{\bibinfo{person}{TripAdvisor}.}
  \bibinfo{year}{2017}\natexlab{}.
\newblock \showarticletitle{Tripadvisor}.
  \bibinfo{address}{\url{https://www.tripadvisor.it/}}.
\newblock


\bibitem[\protect\citeauthoryear{Verma, Ghosh, Saketh, Ganguly, Mitra, and
  Chakraborty}{Verma et~al\mbox{.}}{2018}]%
        {Verma-etal:18}
\bibfield{author}{\bibinfo{person}{Rohit Verma}, \bibinfo{person}{Surjya
  Ghosh}, \bibinfo{person}{Mahankali Saketh}, \bibinfo{person}{Niloy Ganguly},
  \bibinfo{person}{Bivas Mitra}, {and} \bibinfo{person}{Sandip Chakraborty}.}
  \bibinfo{year}{2018}\natexlab{}.
\newblock \showarticletitle{Comfride: a smartphone based system for comfortable
  public transport recommendation}. In \bibinfo{booktitle}{\emph{Proceedings of
  the 12th ACM Conference on Recommender Systems}}
  \emph{(\bibinfo{series}{RecSys '18})}. \bibinfo{publisher}{ACM},
  \bibinfo{address}{New York, NY, USA}, \bibinfo{pages}{181--189}.
\newblock
\showISBNx{978-1-4503-5901-6}
\urldef\tempurl%
\url{https://doi.org/10.1145/3240323.3240359}
\showDOI{\tempurl}


\bibitem[\protect\citeauthoryear{{von Winterfeldt} and Edwards}{{von
  Winterfeldt} and Edwards}{1986}]%
        {vonWinterfeldt:86}
\bibfield{author}{\bibinfo{person}{Detlof {von Winterfeldt}} {and}
  \bibinfo{person}{Ward Edwards}.} \bibinfo{year}{1986}\natexlab{}.
\newblock \bibinfo{booktitle}{\emph{Decision Analysis and Behavioral
  Research}}.
\newblock \bibinfo{publisher}{Cambridge University Press},
  \bibinfo{address}{Cambridge, UK}.
\newblock
\showISBNx{9780521253086}


\bibitem[\protect\citeauthoryear{Wibowo, Siddharthan, Masthoff, and Lin}{Wibowo
  et~al\mbox{.}}{2018}]%
        {Wibowo-etal:18}
\bibfield{author}{\bibinfo{person}{Agung~Toto Wibowo}, \bibinfo{person}{Advaith
  Siddharthan}, \bibinfo{person}{Judith Masthoff}, {and}
  \bibinfo{person}{Chenghua Lin}.} \bibinfo{year}{2018}\natexlab{}.
\newblock \showarticletitle{Incorporating constraints into matrix factorization
  for clothes package recommendation}. In \bibinfo{booktitle}{\emph{Proceedings
  of the 26th Conference on User Modeling, Adaptation and Personalization}}
  \emph{(\bibinfo{series}{UMAP '18})}. \bibinfo{publisher}{ACM},
  \bibinfo{address}{New York, NY, USA}, \bibinfo{pages}{111--119}.
\newblock
\showISBNx{978-1-4503-5589-6}
\urldef\tempurl%
\url{https://doi.org/10.1145/3209219.3209228}
\showDOI{\tempurl}


\bibitem[\protect\citeauthoryear{Zanker, Aschinger, and Jessenitschnig}{Zanker
  et~al\mbox{.}}{2010}]%
        {Zanker-etal:10}
\bibfield{author}{\bibinfo{person}{Markus Zanker}, \bibinfo{person}{Markus
  Aschinger}, {and} \bibinfo{person}{Markus Jessenitschnig}.}
  \bibinfo{year}{2010}\natexlab{}.
\newblock \showarticletitle{Constraint-based personalised configuring of
  product and service bundles}.
\newblock \bibinfo{journal}{\emph{Int. Journal of Mass Customisation}}
  \bibinfo{volume}{3}, \bibinfo{number}{4} (\bibinfo{year}{2010}).
\newblock
\showISSN{1742-4208}
\urldef\tempurl%
\url{https://doi.org/10.1504/IJMASSC.2010.037653}
\showDOI{\tempurl}


\bibitem[\protect\citeauthoryear{Zheng}{Zheng}{2017}]%
        {Zheng:17}
\bibfield{author}{\bibinfo{person}{Yong Zheng}.}
  \bibinfo{year}{2017}\natexlab{}.
\newblock \showarticletitle{Criteria chains: a novel multi-criteria
  recommendation approach}. In \bibinfo{booktitle}{\emph{Proceedings of the
  22nd International Conference on Intelligent User Interfaces}}
  \emph{(\bibinfo{series}{IUI '17})}. \bibinfo{publisher}{ACM},
  \bibinfo{address}{New York, NY, USA}, \bibinfo{pages}{29--33}.
\newblock
\showISBNx{978-1-4503-4348-0}
\urldef\tempurl%
\url{https://doi.org/10.1145/3025171.3025215}
\showDOI{\tempurl}


\end{thebibliography}

\balance

\end{document}